%% file: article.tex
\documentclass[pre,twocolumn,superscriptaddress,floatfix,amssymb,showpacs,aps,10pt]{revtex4-1}
\usepackage{enumerate}
\usepackage{times}
\usepackage{graphicx}
\usepackage{amsmath}
\usepackage{color}
\usepackage{natbib}
\usepackage{array}
\newcolumntype{C}[1]{>{\centering\let\newline\\\arraybackslash\hspace{0pt}}m{#1}}

\usepackage[pdfpagemode=None,colorlinks=true,urlcolor=black,%
linkcolor=blue,citecolor=blue,pdfstartview=FitH]{hyperref}
\bibpunct{[}{]}{,}{n}{}{;}
\usepackage{lipsum}

\graphicspath{{fig/}} 
\begin{document}

\input{copyrightaps.tex}

\title{Chaperone-assisted translocation of flexible polymers in three dimensions}

\author{P. M. Suhonen}
\author{R. P. Linna}
\email{Corresponding author: riku.linna@aalto.fi}
\affiliation{Department of Computer Science, Aalto University, P.O. Box 15400, FI-00076 Aalto, Finland}


\begin{abstract}
Polymer translocation through a nanometer-scale pore assisted by chaperones binding to the polymer is a process encountered {\it in vivo} for proteins. Studying the relevant models by computer simulations is computationally demanding. Accordingly, previous studies are either for stiff polymers in three dimensions or flexible polymers in two dimensions. Here, we study chaperone-assisted translocation of flexible polymers in three dimensions using Langevin dynamics. We show that differences in binding mechanisms, more specifically, whether a chaperone can bind to a single or multiple sites on the polymer, lead to substantial differences in translocation dynamics in three dimensions. We show that the single-binding mode leads to dynamics that is very much like that in the constant-force driven translocation and accordingly mainly determined by tension propagation on the {\it cis} side. We obtain $\beta \approx 1.26$ for the exponent for the scaling of the translocation time with polymer length. This fairly low value can be explained by the additional friction due to binding particles. The multiple-site binding leads to translocation whose dynamics is mainly determined by the {\it trans} side. For this process we obtain $\beta \approx 1.36$. This value can be explained by our derivation of $\beta = 4/3$ for constant-bias translocation, where translocated polymer segments form a globule on the {\it trans} side. Our results pave the way for understanding and utilizing chaperone-assisted translocation where variations in microscopic details lead to rich variations in the emerging dynamics.
\end{abstract}

\maketitle
\section{Introduction}\label{sec:intro}
Polymer translocation through a nanopore has been a topic of major interest ever since Kasianowicz et al. suggested that the process could be used for inexpensive and fast DNA sequencing~\cite{Kasianowicz96}. There is a plethora of studies to explain different aspects of the process in various circumstances. For a recent review, see~\cite{Palyulin14}.

Among different variants of polymer translocation, the process driven by binding particles (BiPs) has gotten less attention. In this form of polymer translocation, freely diffusing BiPs bind to the translocating polymer on the \textit{trans} side. The bound BiPs block the polymer from reentering the pore and hence prevent its backwards motion towards the \textit{cis} side. This Brownian ratcheting mechanism creates a bias to the polymer's diffusion and drives the translocation. 

An example in cell biology of a similar process is the protein translocation into the lumen of endoplasmic reticulum and into the mitochondrial matrix~\cite{Matlack99,Alberts02,Neupert02,Neupert07}. It is believed that during the translocation, auxiliary proteins called chaperones bind to the translocating polypeptide chain, which causes Brownian ratcheting. 

The Brownian ratcheting was first theoretically studied in Ref.~\cite{Simon92}. After this the topic has been discussed in a number of publications, see {\it e.g.}~\cite{Peskin93,Sung96,Elston00,Liebermeister01,Elston02,Farkas03,Zandi03,Ambjornsson04,Kafri04,Ambjornsson05,DeLosRios06,Inamdar06,DOrsogna07,Abdolvahab08,Metzler10,Depperschmidt10,Abdolvahab11,Abdolvahab11b,Yu11,Yu12,Depperschmidt13,Yu14,Adhikari15}, some of which are computational studies. Monte Carlo simulations of the process have been reported in Refs.~\cite{Elston02,DOrsogna07,Abdolvahab08,Abdolvahab11,Abdolvahab11b}. Extra care has to be taken to make sure Monte Carlo simulations capture the correct dynamics of translocation processes~\cite{Lehtola08}. In this respect, Langevin dynamics (LD) simulations are a more straight forward approach~\cite{Farkas03,Zandi03,Yu11,Yu12,Yu14,Adhikari15}. Presumably due to the heavy computational requirements, the three dimensional BiP driven translocation has not been much investigated by LD. The systems studied by LD are fairly small or in two dimensions. To our knowledge the only existing three-dimensional study concerns BiPs driving stiff chains~\cite{Zandi03}, which does not capture the true dynamics of non-rigid polymers.

The motivation for studying binding-particle driven translocation of stiff polymers was to facilitate a theoretical basis for the more complicated case of flexible polymers~\cite{Zandi03}. It was argued that the essential features of the process would be covered by including the dynamics within the polymer's persistence length from the pore. However, from the numerous studies of translocation driven by force applied at the pore we know  that changes in the conformation of the non-rigid polymer during translocation largely determines the dynamics, see {\it e.g.}~\cite{Sakaue07,Lehtola09,Rowghanian11}. 

Flexible BiP-driven polymers and the effect of flexibility has been studied in two dimensions~\cite{Yu11,Yu14,Adhikari15}. With stiff polymers particle binding was unambiguous: BiPs can only bind to one site (polymer segment) at a time. Introducing flexibility changes this. In \cite{Yu11} and \cite{Yu14} BiPs were allowed to bind to multiple polymer segments simultaneously. However, in many known cases in cellular biology a protein has only a single binding site for interactions with another molecule. When this is the case, a binding model that restricts the binding of BiPs to only a single segment of the polymer at a time should be used.

Here, we investigate the BiP driven translocation of flexible polymers in three dimensions using Langevin dynamics. We apply two different binding models. In the one-to-one (OTO) binding model we restrict the binding of BiPs to only a single monomer at a time. In the all-to-all (ATA) binding model we allow BiPs to bind to all monomers in their vicinity. Regarding previous studies the OTO model introduces polymer flexibility to single site binding of stiff polymers, whereas the ATA model introduces the third dimension to the two-dimensional models studied in~\cite{Yu11,Adhikari15}.

We show that the processes in three dimensions are crucially different from the processes in two dimensions and that changing the binding mechanism completely changes the process in three dimensions. We compare translocation driven by OTO binding to translocation driven by a constant pore force and show that also the dynamics of the OTO driven process is mainly determined by tension propagation in the polymer segment on the {\it cis} side. Close resemblance in the tension propagation of the BiP-assisted and pore force driven translocation was recently seen in two dimensions~\cite{Adhikari15}.

In what follows, we first outline the computational setting by describing models used for polymers, binding particles, dynamics, and the pore and the membrane. We then report and analyze the results from our simulations. Finally, we recap the main conclusions of our study.

\section{The computational models}\label{sec:sm}

\begin{figure}[b]
\includegraphics[width=0.80\linewidth]{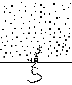}
\caption{Depiction of the simulation setup of a polymer undergoing binding particle (BiP) driven translocation from the \textit{cis} side (bottom) to the \textit{trans} side (top). Polymer beads (PB) are drawn as circles and BiPs as squares. The \textit{cis} and \textit{trans} sides are separated by a slip-wall membrane of thickness $3\sigma$. In the membrane there is a pore of diameter $2\sigma$ that allows the polymer to pass through. To prevent BiPs from diffusing away in the x- and y-directions, periodic boundary conditions are applied. For the z-direction on the \textit{trans} side, diffusion is prevented by a slip-wall sufficiently far away from the pore.}
\label{fig:SimulationSetup}
\end{figure}

The three-dimensional simulation space consisting of a translocating polymer, binding particles (BiPs), membrane walls and periodic boundaries is depicted in Fig.~\ref{fig:SimulationSetup}. In what follows \textit{cis} and \textit{trans}  signify the sides of the membrane on which the polymer resides initially and to which it translocates, respectively.

\subsection{The polymer model}\label{sec:pm}

Excluded volume interactions of the polymer are taken into account via a Lennard-Jones(L-J) potential acting between any two polymer beads (PB),
\begin{align}\label{equ:LJ}
U_{LJ} = 4 \epsilon \left[ \left(\frac{\sigma}{r}\right)^{12} - \left(\frac{\sigma}{r}\right)^{6} + \frac{1}{4}\right], r \leq 2^{1/6}\sigma.
\end{align}
Here, $\epsilon=1.0$ is the strength of the interaction, $\sigma=1.0$ the length scale of the interaction, and $r$ the current distance between two PBs. By setting the cut-off distance $r=2^{1/6}\sigma$ to exclude attractive interactions we model the polymer to be immersed in good solvent.

The polymer is modeled as a freely-jointed bead spring chain. Adjacent PBs are connected together by a finitely extensible nonlinear elastic (FENE) potential
\begin{align}\label{equ:FENE}
U_F = -\frac{K}{2}R^2 \ln{\left(1-\frac{r^2}{R^2}\right)},
\end{align}
where $K = \frac{30}{\sigma^2}$ is the strength of the attractive interaction, $R = 1.5 \sigma$ is the maximum distance, and $r$ the current distance between two connected PBs.

\subsection{The binding particle model}\label{sec:cm}
The interaction between any two BiPs is modeled with the repulsive L-J interaction of Eq.~\eqref{equ:LJ}. A slightly different L-J potential is used for modeling the interaction between a BiP and a PB. First, a PB and a BiP can bind together via the attractive part of the L-J interaction. Second, we use $\epsilon_b$ instead of $\epsilon$ for the binding strength of the BiPs to the PBs. The BiP-PB interaction hence takes the form

\begin{align}\label{equ:LJBind}
U_{LJ} = 4 \epsilon_b \left[ \left(\frac{\sigma}{r}\right)^{12} - \left(\frac{\sigma}{r}\right)^{6} + \frac{1}{4}\right], r \leq r_{max}
\end{align}

The binding strength is chosen to be $\epsilon_b=8.0$. Only when investigating the effect of the binding strength $\epsilon_b$ is varied between $1.0$ and $64.0$. 

The binding and unbinding is controlled via the threshold distance $r_{max}$. We conduct our simulations with two different models for binding. In both models the binding is described by Eq.~\eqref{equ:LJBind} and can only take place between a BiP and a PB. In the ATA binding model every BiP-PB pair can bind together when they are within the distance $r_{bind}=1.84\sigma$ of each other. Eq.~\eqref{equ:LJBind} is hence used with $r_{max}=r_{bind}$ for all BiP-PB pairs. This allows each BiP/PB to bind to many PBs/BiPs simultaneously. ATA hence corresponds to the inter-segment binding model of Refs.~\cite{Yu11}~and~\cite{Yu14}.
In contrast, in OTO binding model each BiP is allowed to bind to only one PB at a time. When an unbound BiP and an unbound PB are within $r_{bind}$ of each other, a binding takes place and Eq.~\eqref{equ:LJBind} is used with $r_{max}=r_{bind}$. For any BiP/PB interacting with an already bound PB/BiP $r_{max}=2^{1/6}\sigma$ and only the repulsive part is applied. A BiP-PB pair is considered broken if the BiP and PB get farther than $r_{bind}$ apart.

For both models binding between a PB and a BiP can only occur if the PB has entered the \textit{trans} side. This prevents binding of a BiP to a PB that is still inside the pore and, consequently, the BiP from pulling the PB from the pore to the {\it trans} side. If a bound PB re-enters the pore, its binding to the BiP is not broken.

Since the binding and unbinding take place according to the distance of the BiP and the monomer to which it binds, the stochastic nature of this process comes about via the stochastic motion of the particles. Adding explicit binding and unbinding rates would give more freedom in defining {\it e.g.} highly asymmetrical binding and unbinding probabilities. This would, however, slow down the translocation process and make the simulation of the three-dimensional chaperon-assisted translocation computationally an overwhelming task. The binding/unbinding described here is used in the previous studies in two dimensions, which enables us to make direct comparisons to them. 

\subsection{The dynamics of polymer and BiPs}\label{sec:dynamics}

The dynamics for the point-like PB and BiP particles is implemented using Ermak's version of Langevin dynamics~\cite{Ermak80}. The Langevin equation governing the dynamics of a particle indexed $i$ is written as
\begin{align}\label{equ:langevin}
\dot{\textbf{p}}_i=-\xi\textbf{p}_i+\pmb{\eta}_i(t)+\textbf{f}_i(\textbf{r}_i),
\end{align}
where $\textbf{p}_i$ is the momentum of the particle and $\dot{\textbf{p}}_i$ its time derivative, $\xi$ is the friction coefficient of the implicit solvent, $\pmb{\eta}_i$ the resultant random force exerted on the particle, $\textbf{f}_i(\textbf{r}_i)$ the resultant force exerted on the particle, and $\textbf{r}_i$ the position of the particle. The velocity Verlet algorithm is used to integrate the positions and velocities of the particles related by the Langevin equation~\cite{vanGunsteren77}. 

Parameter values used in the simulations are given in reduced units. The Boltzmann constant  $k_B=1.0$ and the temperature $T=1.0$. The time step $\delta t =0.001$ and the friction coefficient $\xi=0.5$ to which we also relate $\pmb{\eta}_i(t)$ according to the fluctuation dissipation theorem. The masses of both PBs and BiPs are $m=16.0$.

\subsection{The pore, membrane, and boundary conditions}

The simulation space consists of two compartments separated by a membrane. The membrane is modeled by a wall of thickness $3\sigma$. Slip boundary conditions are applied for all beads colliding the two wall surfaces. A circular pore of diameter $2\sigma$ penetrates the wall allowing PBs to pass from one side to the other. BiPs residing on the \textit{trans} side cannot enter the pore.

The pore is implemented by a harmonic force that pulls the PBs toward an axis orthogonal to the wall surfaces
\begin{align}\label{equ:PoreHarmonic}
f_h = -k_p r_p - c v_p.
\end{align}
Here, $r_p$ is the distance of the PB from the pore axis and $v_p$ is the velocity of the PB perpendicular to the pore axis. The coefficient values were chosen as $k_p=100.0$ and $c=1.0$. In addition to the harmonic force aligning the polymer, hairpinning is prevented also by only allowing PBs to enter the pore sequentially.

Periodic boundaries in x- and y-directions and a slip-wall perpendicular to the z-direction prevent the BiPs on the \textit{trans} side from diffusing away. The periodic boundary conditions and the wall are applied for both BiPs and PBs. The slip-wall in the z-direction is placed so far that only few of the longest ($N=400$) polymers under OTO binding touch the wall.

\subsection{About the simulations}\label{sec:simulations}

At the start of all simulations almost the entire polymer is on the \textit{cis} side. A short segment is inside the pore and two monomers in the head protrude to the \textit{trans} side. All the BiPs are on the \textit{trans} side. See the first snapshot in Fig.~\ref{fig:SimPicsAll}.

Simulations are started from equilibrated polymer conformations. A polymer is equilibrated while keeping the polymer end fixed. During equilibration we measure the radius of gyration $R_g^2 = \sum_{i=1}^N (\textbf{r}_i - \textbf{r}_{cm})^2$, where $\textbf{r}_{cm}$ is the polymer's center of mass. An equilibrium conformation is considered to be reached when the time-averaged $R_g$ has converged to a stable value. After the polymer equilibration the BiPs on the \textit{trans} side are also let to find an equilibrium distribution and bind to the two PBs on the \textit{trans} side. After this the polymer is released, and translocation begins.

For all sets of presented data we have conducted $300$ simulations. There are some exceptions: For polymers of length $N=400$ with \textit{cis} dynamics excluded, we conducted $100$ simulations each. For simulations used to calculate the equilibrium $R_g$, $50$ time-averaged simulations were used. It should also be noted that for small binding strengths $\epsilon_b$ a number of translocations do not complete due to some polymers sliding back to the \textit{cis} side. In these cases the number of simulations can be considerably less than $300$. For the intermediate value $\epsilon_b=8$ used in most of our simulations around $10\%$ of the polymers do not translocate.

In the simulations we fix the concentration of free BiPs to $c_f=1/40$ unless stated otherwise. The exceptions are simulations for investigating the effect of $c_f$. Here $c_f$ is chosen between $1/320$ and $1/5$. The value of $c_f$ is maintaned by creating a new BiP at the edge of the simulation space if $c_f$ drops below a threshold value.

\section{Results}\label{sec:res}

\subsection{Different binding causes visible difference in  transclocation}\label{sec:ovbind}

\begin{figure}[t]
\includegraphics[width=0.32\linewidth]{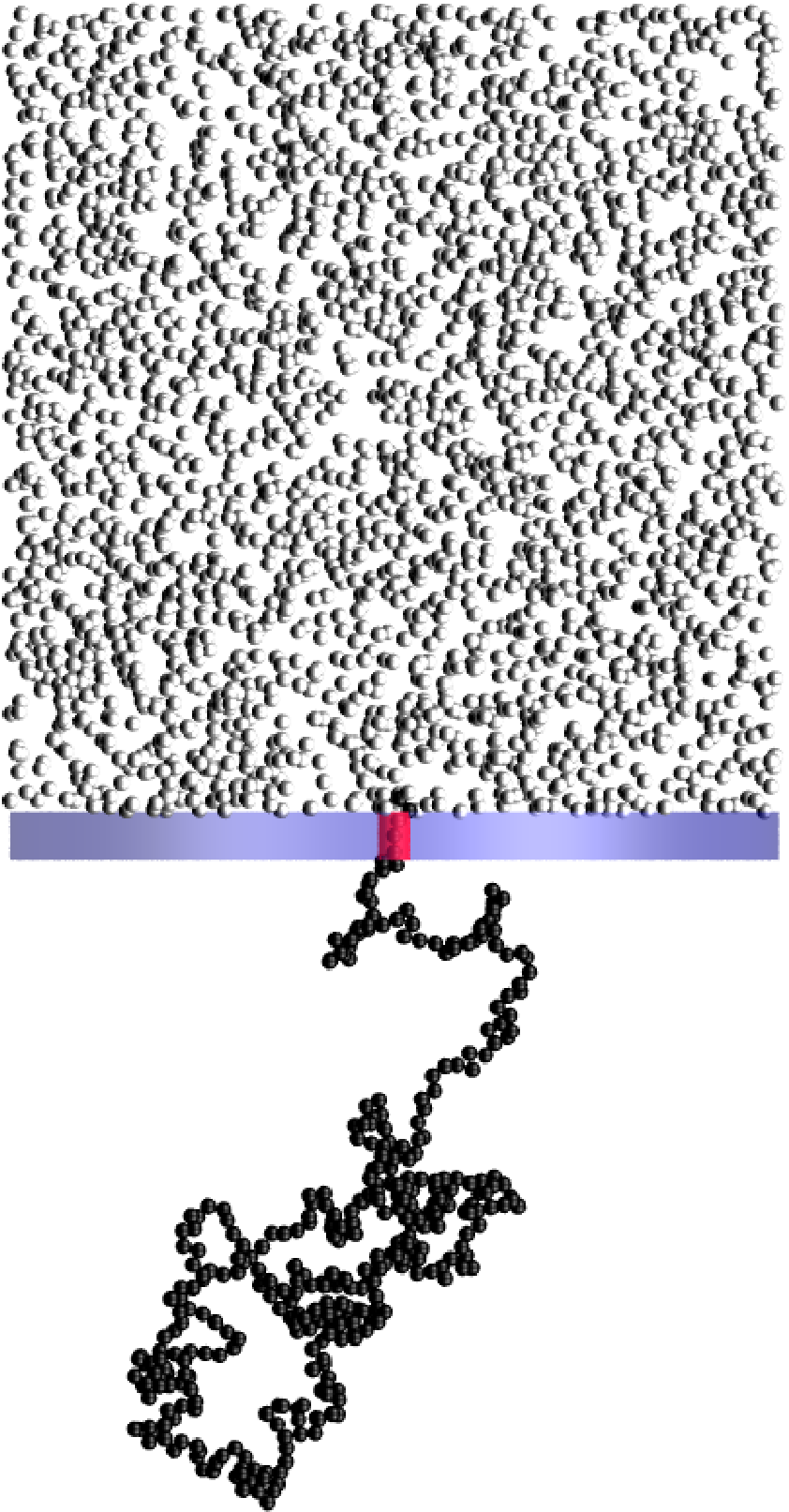}
\includegraphics[width=0.32\linewidth]{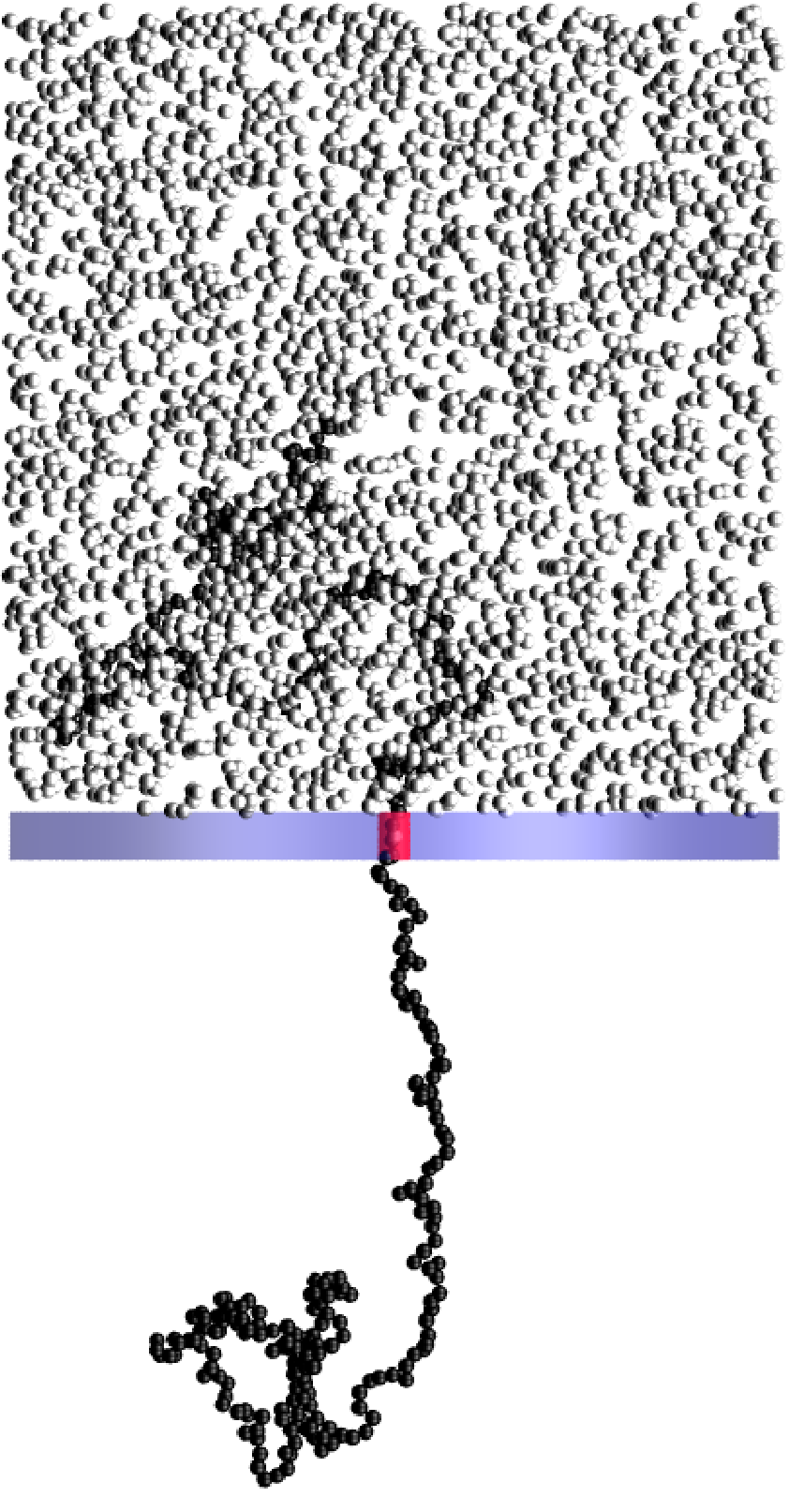}
\includegraphics[width=0.32\linewidth]{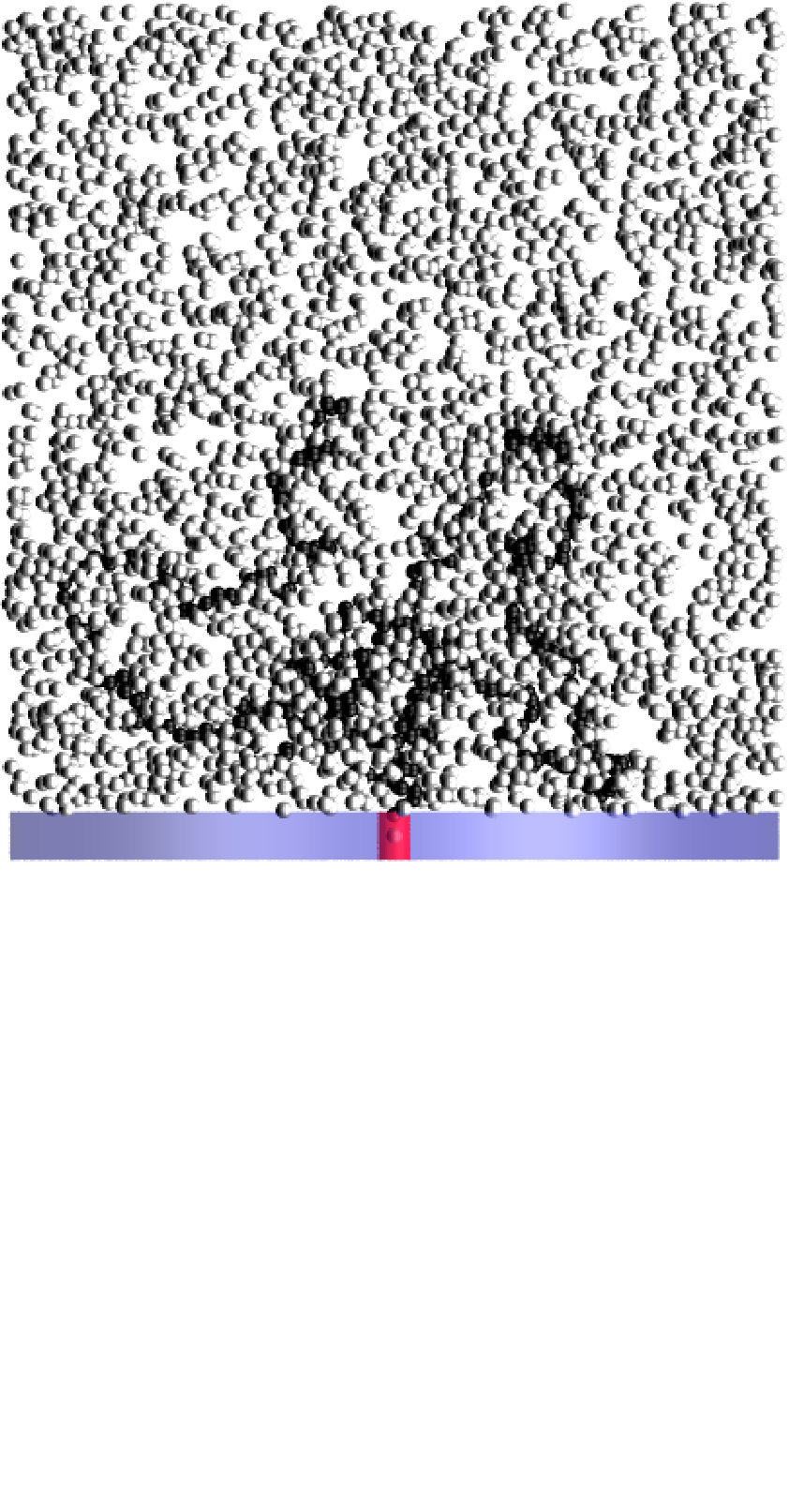}
\\
\includegraphics[width=0.32\linewidth]{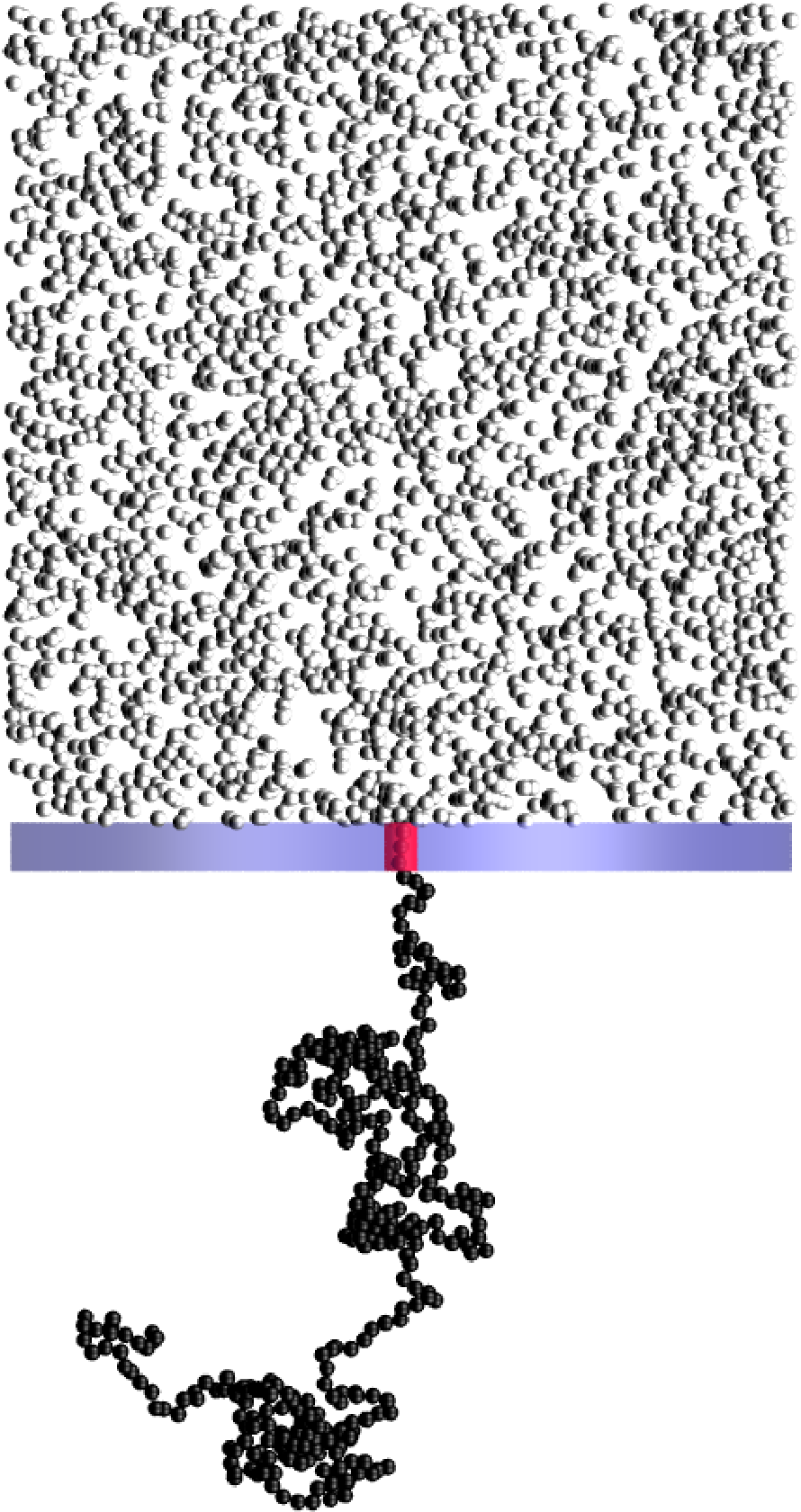}
\includegraphics[width=0.32\linewidth]{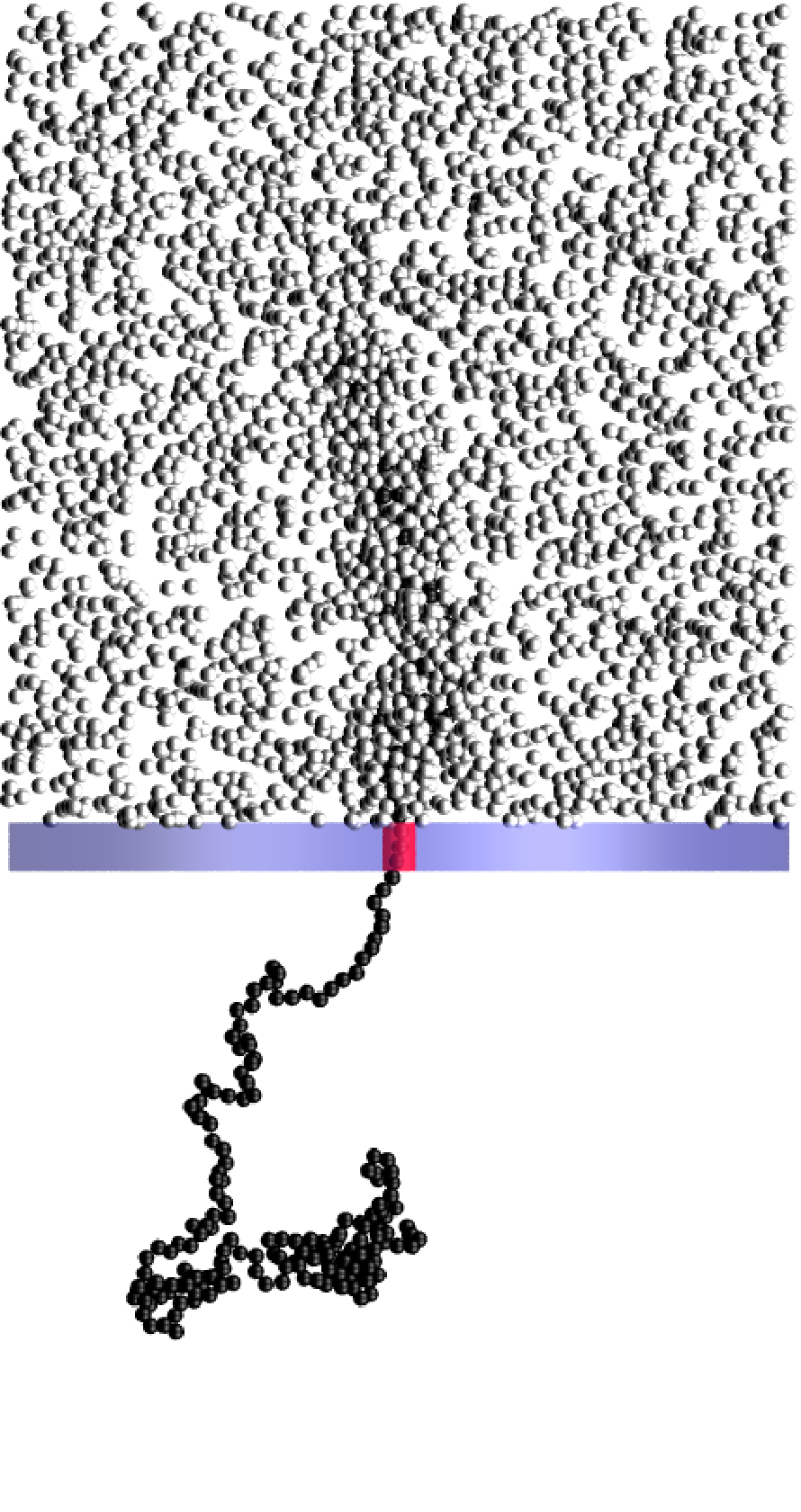}
\includegraphics[width=0.32\linewidth]{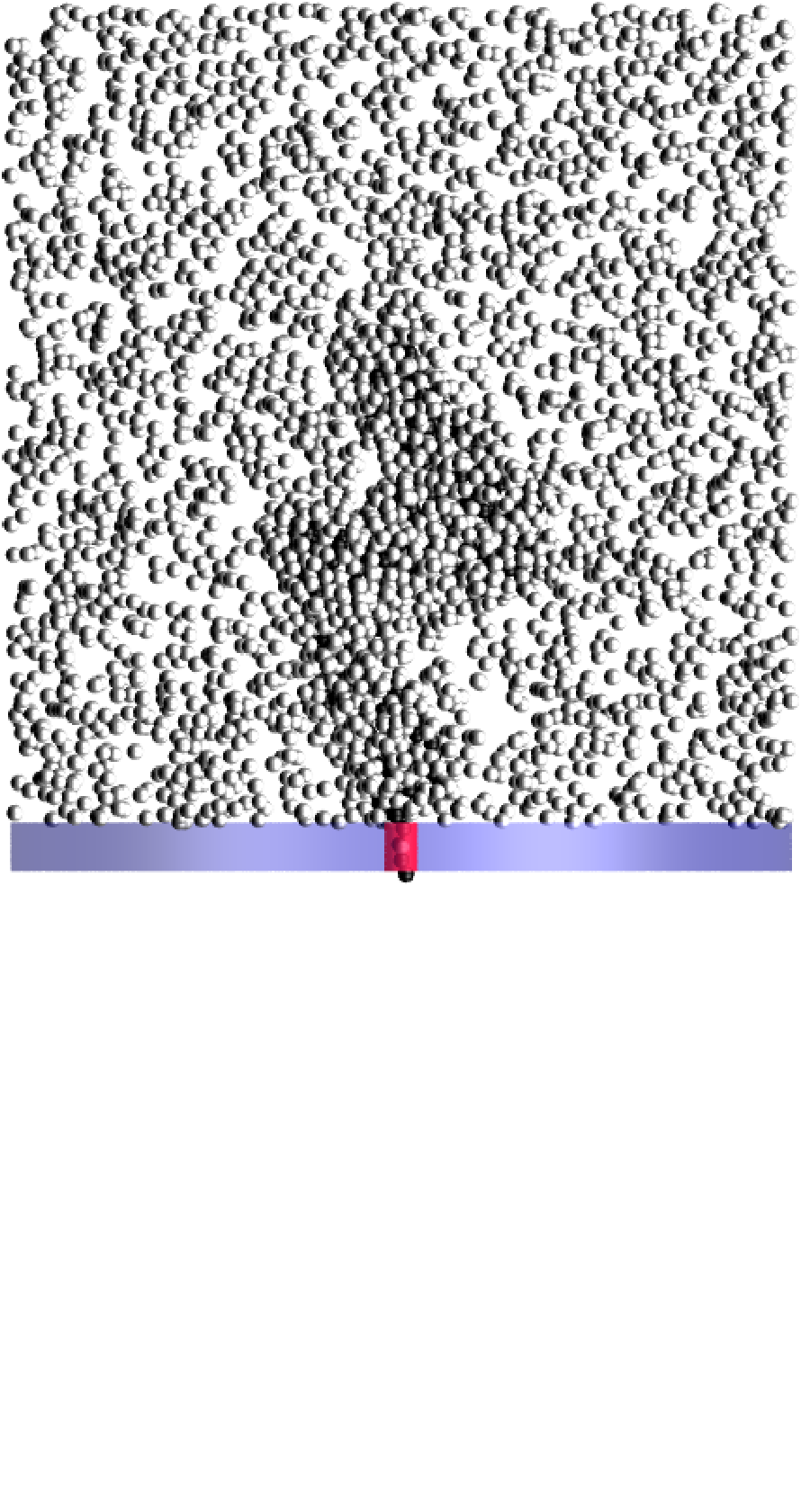}
\caption{(Color online) Snapshots from simulations of BiP driven translocation using OTO, upper row, and ATA, lower row. The leftmost snapshots are taken at the start of the simulations, the center snapshots when half of the polymer has translocated, and the rightmost snapshots at the end of the process.}
\label{fig:SimPicsAll}
\end{figure}

\begin{figure}[t]
\includegraphics[width=0.8\linewidth]{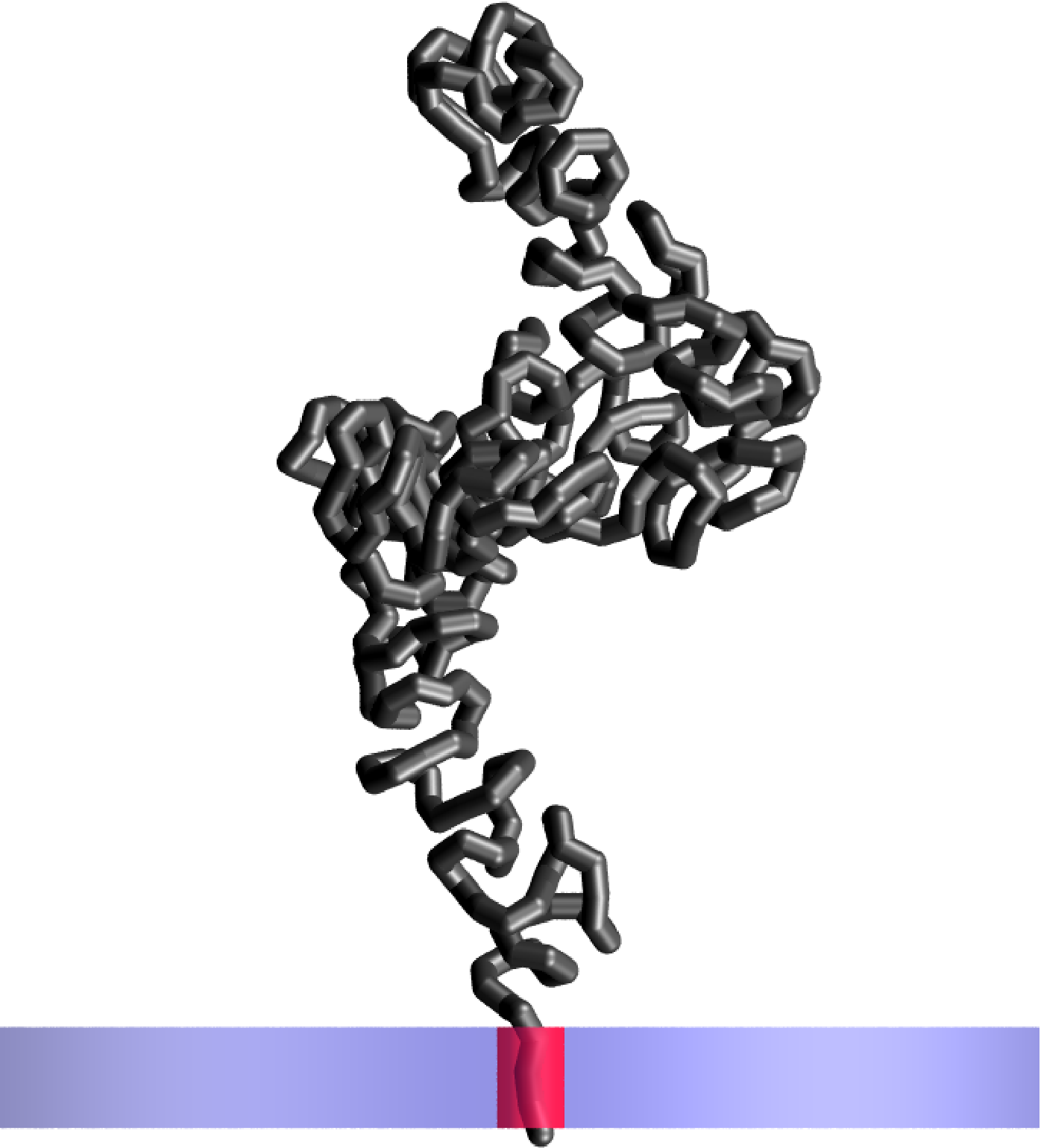}
\caption{(Color online) A simplified snapshot from the end conformation of the ATA simulation of Fig.~\ref{fig:SimPicsAll}. The BiPs have been omitted to show how the polymer coils around itself forming helical segments.}
\label{fig:SimPicsAllToAllHelicality}
\end{figure}

The snapshots from the simulations in Fig.~\ref{fig:SimPicsAll} show how the two different binding models affect translocation. They are taken from simulations conducted with polymers of length $N=400$. The snapshots in the upper and lower rows are from single simulations using OTO and ATA, respectively. In OTO the polymer takes a diffuse conformation on the \textit{trans} side, whereas ATA brings the polymer to a highly folded conformation consisting of helical regions, see Fig.~\ref{fig:SimPicsAllToAllHelicality}. The strong folding markedly differs the translocation driven by ATA binding in $3$ dimensions from the corresponding process in $2$ dimensions. In $2$D the intersegmental binding in ATA binding is much more restricted than in $3$D, so the difference to translocation driven by OTO binding is not as significant in $2$D as in $3$D.

\subsection{Relaxation of the polymer segment on the \textit{trans} side}\label{sec:trans}

In our previous studies on driven polymer translocation we measured $R_g$ for segments on the {\it trans} side to determine if the translocation of segments was faster than relaxation of translocated segments to equilibrium. We found that translocated segments do not have time to relax but are driven increasingly further out of equilibrium as the number of translocated monomers $N_{tr}$ increases. This shows as the difference $R_g^{eq}(N_{tr})-R_g(N_{tr})$ increasing with $N_{tr}$, where $R_g^{eq}(N_{tr})$ is the radius of gyration for an equilibrium conformation of a polymer of length $N_{tr}$~\cite{Lehtola09,Suhonen14}.

\begin{figure}[t]
\includegraphics[width=1.0\linewidth]{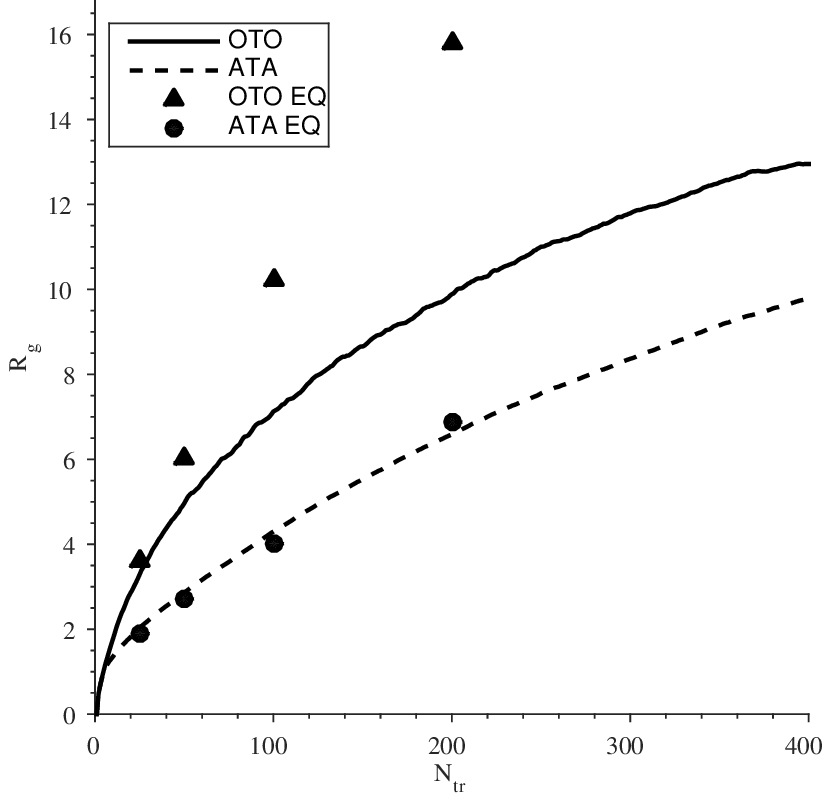}
\caption{$R_g$ of the \textit{trans} side polymer segment as a function of the number of translocated monomers $N_{tr}$ during the BiP-driven translocation for both OTO and ATA models. Also $R_g$ for equilibrated polymers, $R_g^{eq}$, of equal lengths are shown for comparison (triangles and circles).}
\label{fig:TransRG}
\end{figure}

We apply the same method here. Fig.~\ref{fig:TransRG} shows how $R_g(N_{tr})$ for OTO and ATA evolve during translocation. $R_g^{eq}(N_{tr})$ for both models at the same BiP concentration $c_f$ is also shown. $R_g(N_{tr})$ for OTO is seen to be much larger than for ATA as expected due to the polymer in ATA partially folding, see Fig.~\ref{fig:SimPicsAll}. Still, $R_g$ for OTO is much smaller than the corresponding $R_g^{eq}$ indicating that although the process is driven by incomplete Brownian ratcheting, the \textit{trans} side polymer segment is driven out of equilibrium. In contrast, the {\it trans} side $R_g$ of the polymer in the ATA model follows $R_g^{eq}$.

\subsection{Waiting times: contribution of tension propagation}\label{sec:tensionpropagation}

Waiting time $t_w(s)$ is the average time for the bead $s$ to exit the pore after the bead $s+1$ has exited. Its measurement is the most straightforward way to gain understanding on translocation dynamics. We calculate waiting times by subtracting the last passage time of the current bead from that of the previous bead. We have checked that using first passage times instead does not change the  waiting time profiles.

In order to asses the role of the {\it cis} side on the dynamics of the BiP-driven translocation models we also simulate a modified model where the polymer beads on the \textit{cis} side are excluded. In this modified model we do not have a polymer segment on the {\it cis} side but generate PBs at the pore entrance. Should the polymer slide back, the PBs entering the \textit{cis} side are removed from the polymer. We have previously used this method in connection with the driven polymer translocation~\cite{Suhonen14}.

\begin{figure}[t]
\includegraphics[width=0.49\linewidth]{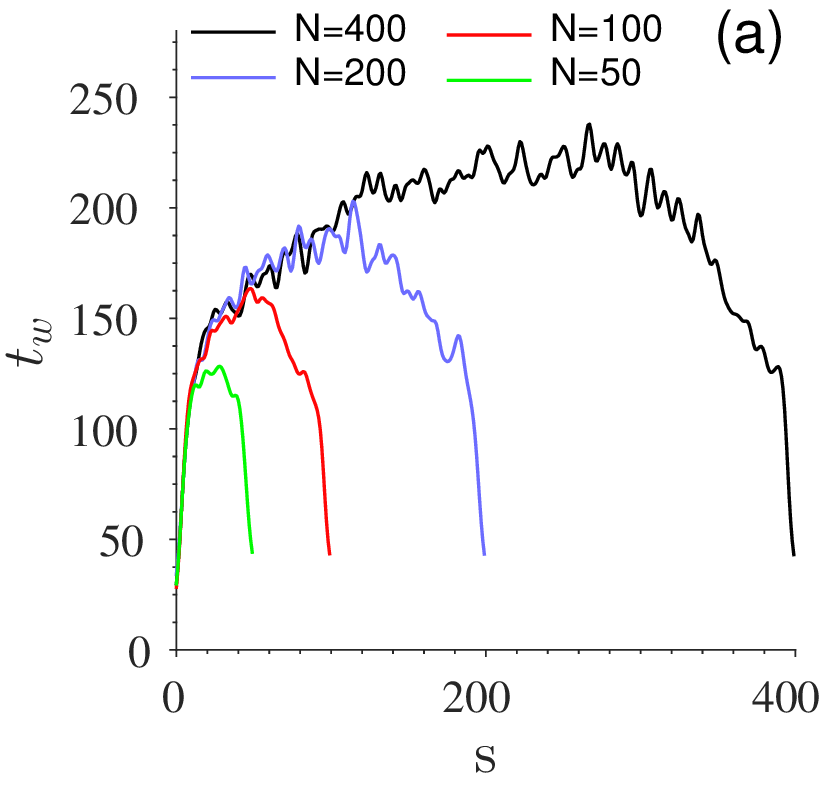}
\includegraphics[width=0.49\linewidth]{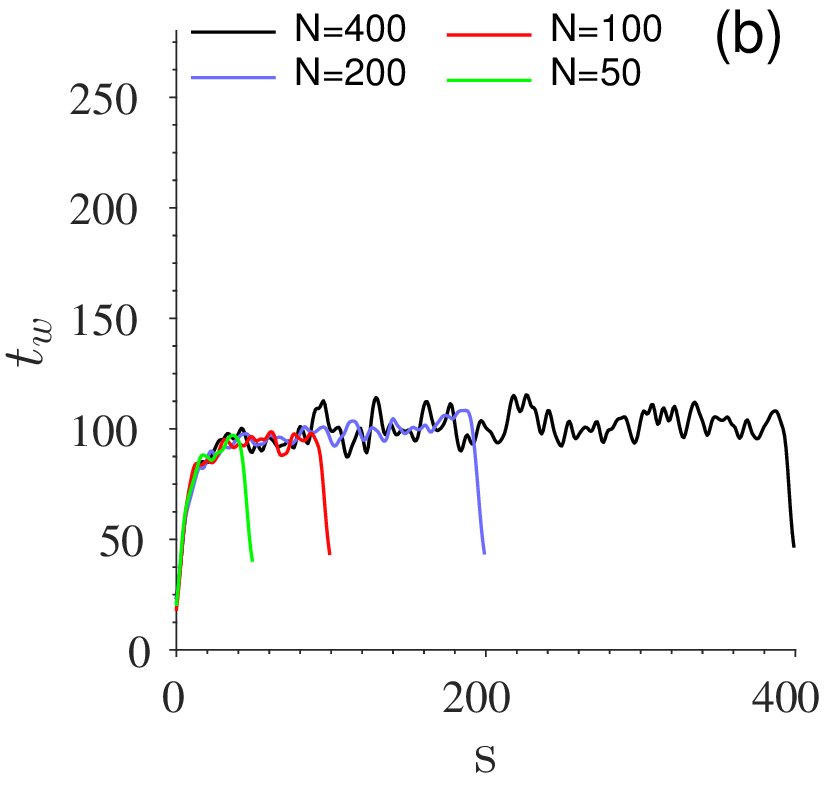}
\includegraphics[width=0.49\linewidth]{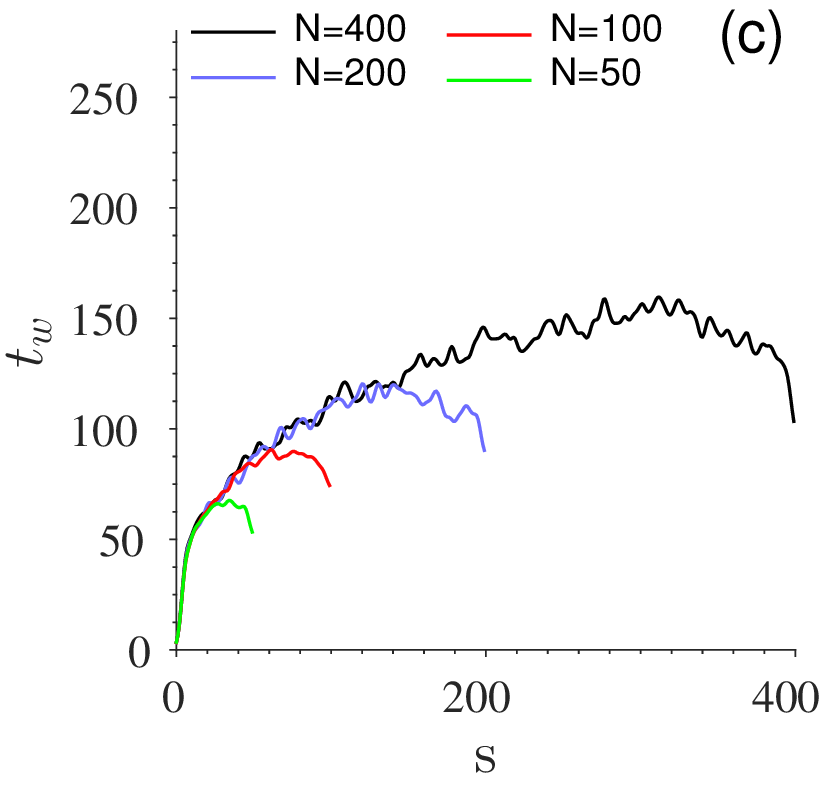}
\includegraphics[width=0.49\linewidth]{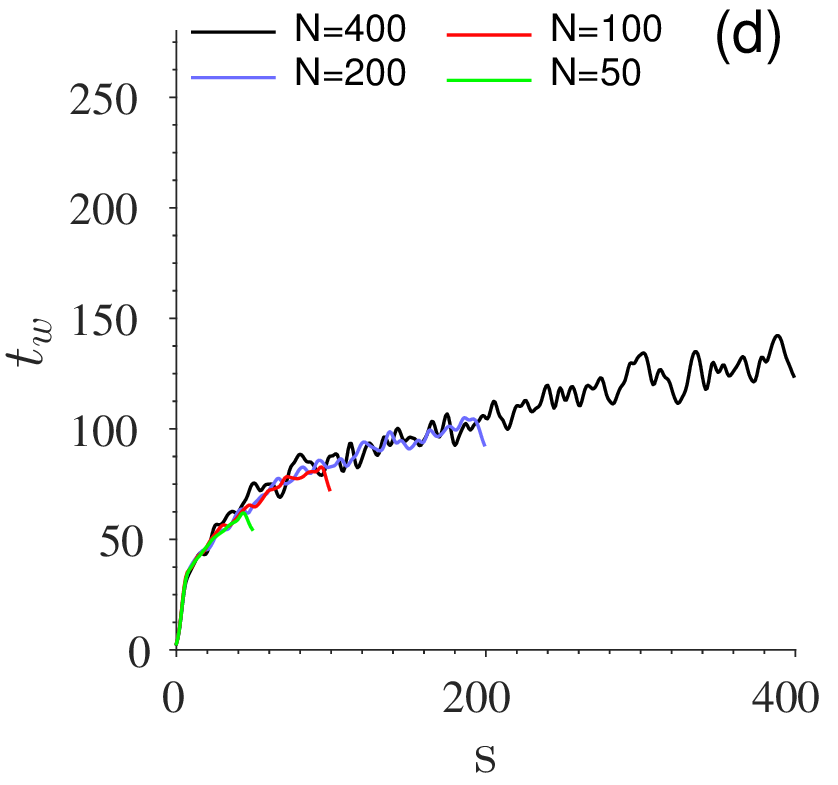}
\caption{(Color online) Waiting times for OTO: (a) full  model (b) models where the contribution from the {\it cis} side is excluded; for ATA: (c) full  model (d) models where the contribution from the {\it cis} side is excluded. Polymer lengths $N=50$, $100$, $200$, and $400$.}
\label{fig:WaitingTimes}
\end{figure}

Fig.~\ref{fig:WaitingTimes} shows the ensemble averages of the waiting times $t_w(s)$ for the full (a) and modified (b) OTO model and for the full (c) and modified (d) ATA model. The waiting time data is inherently noisy. The amount of statistics required to suppress the noise to an insignificant level would be unfeasible for the system sizes used here. Accordingly, the presented data has been slightly Gaussian filtered for improved clarity.

ATA binding induces stronger bias than OTO binding, so the waiting times for ATA are clearly shorter. Also the shapes of the waiting time profiles for ATA and OTO are clearly different.

Excluding the {\it cis} side dynamics has a dramatically different effect on ATA and OTO models. The waiting time profile for OTO becomes almost flat when the {\it cis} side is excluded, whereas the $t_w(s)$ for ATA change only mildly. The stronger binding on the {\it trans} side in the ATA model not only speeds up the translocation but also enhances the correlations along the polymer on the {\it trans} side, as seen in Figs.~\ref{fig:SimPicsAll} and \ref{fig:SimPicsAllToAllHelicality}. Accordingly, in the ATA binding the friction for the movement of the polymer segment on the {\it trans} side is larger than in the OTO binding. Consequently, the {\it trans} side has a more dominating role in the translocation dynamics of ATA. It is in place to note here that the larger bias of the ATA model more than compensates for this larger {\it trans} side friction compared to the OTO model.

The contribution from the {\it cis} side comes from the initial conformation and the tension propagating along the polymer contour. Like in all processes where a polymer from an unconstrained conformation is driven by some means through a pore the dynamics is subdiffusive. For the subdiffusive motion the dominant {\it cis} side contribution is expected to be tension propagation, as found for the driven translocation~\cite{Sakaue07,Lehtola09,Rowghanian11,Ikonen12}. In our simulations the dynamics for OTO binding is dominantly determined by the {\it cis} side, see Figs.~\ref{fig:WaitingTimes} (a) and (b). Hence, we expect tension propagation to play a significant role in the dynamics for OTO.

\begin{figure}[t]
\includegraphics[width=0.6\linewidth]{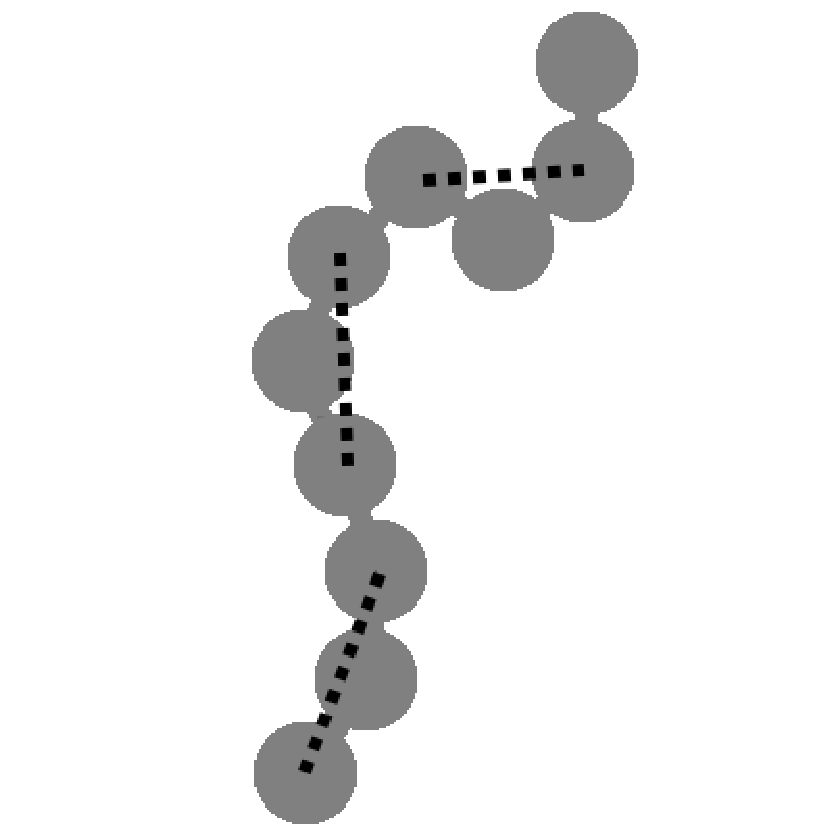}
\caption{In the course of the translocation all two-bond distances between PBs (exemplified by the dashed lines) are computed to quantify polymer's straightening. A longer two-bond distance indicates straighter polymer and stronger tension.}
\label{fig:2BondDist}
\end{figure}

To track the tension propagation during translocation, we apply the same measure for polymer straightening that we successfully used in connection with driven polymer translocation~\cite{Suhonen14}. We measure the distance between all two beads separated by two bonds along the polymer chain for each discrete value of the translocation coordinate $s$, see Fig.~\ref{fig:2BondDist}. For a more detailed description of the measurement of tension during translocation see~\cite{Suhonen14}.

\begin{figure}[t]
\includegraphics[width=0.49\linewidth]{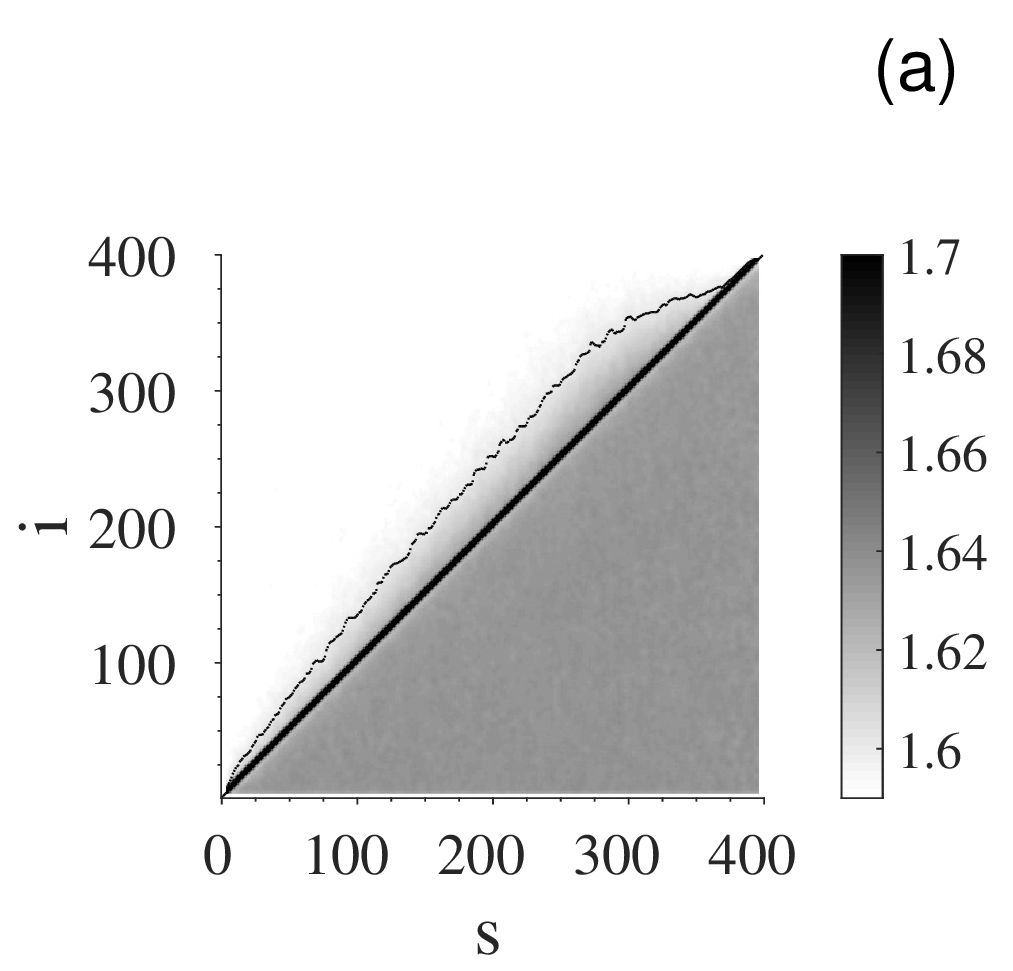}
\includegraphics[width=0.49\linewidth]{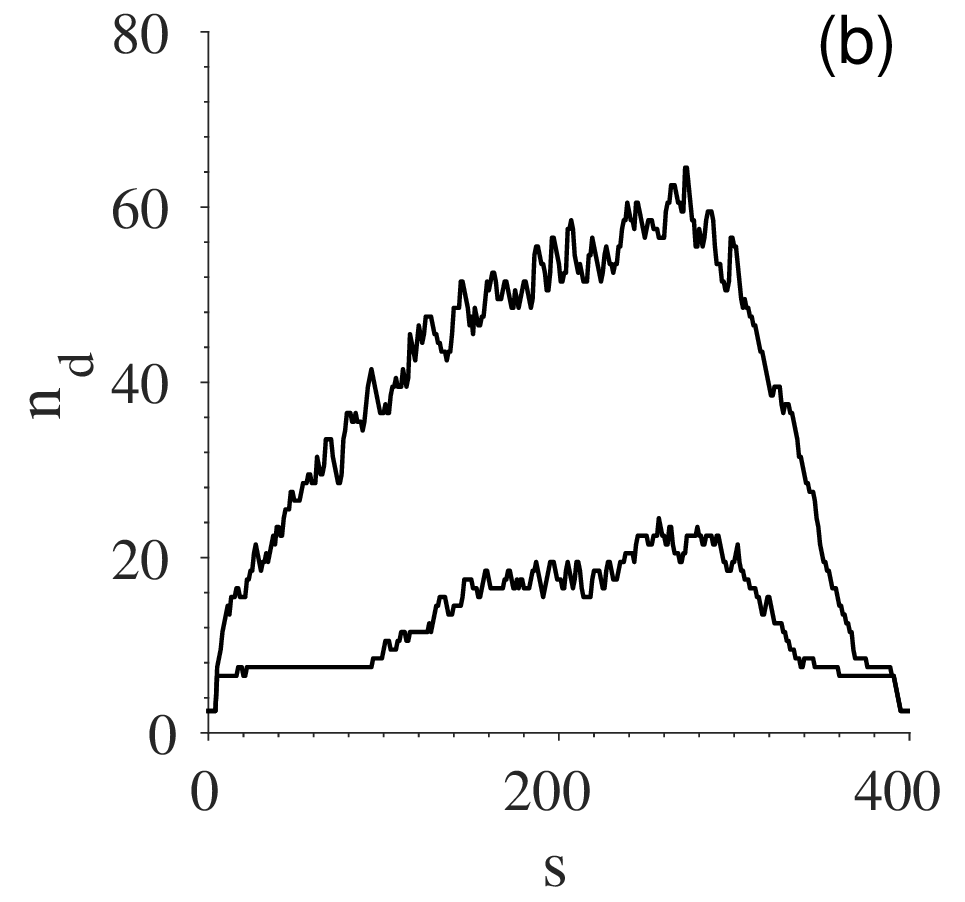}\\
\includegraphics[width=0.49\linewidth]{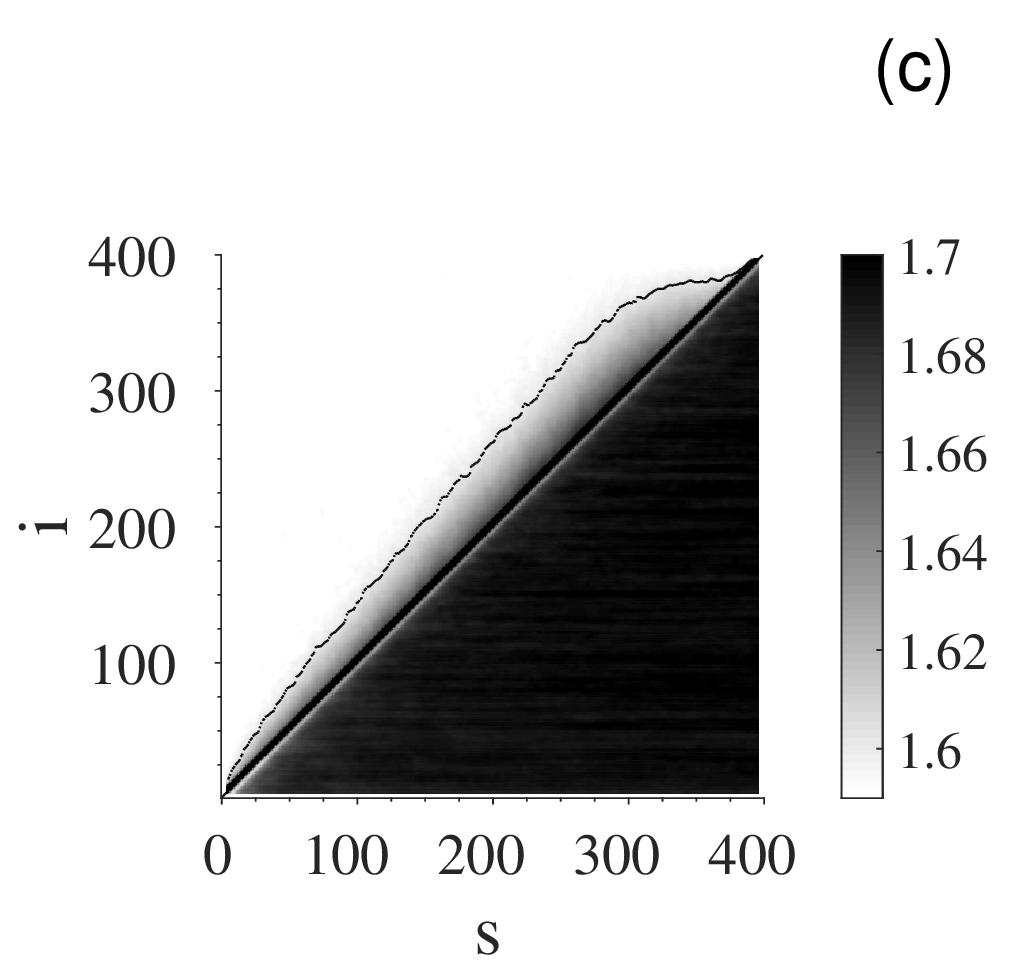}
\includegraphics[width=0.49\linewidth]{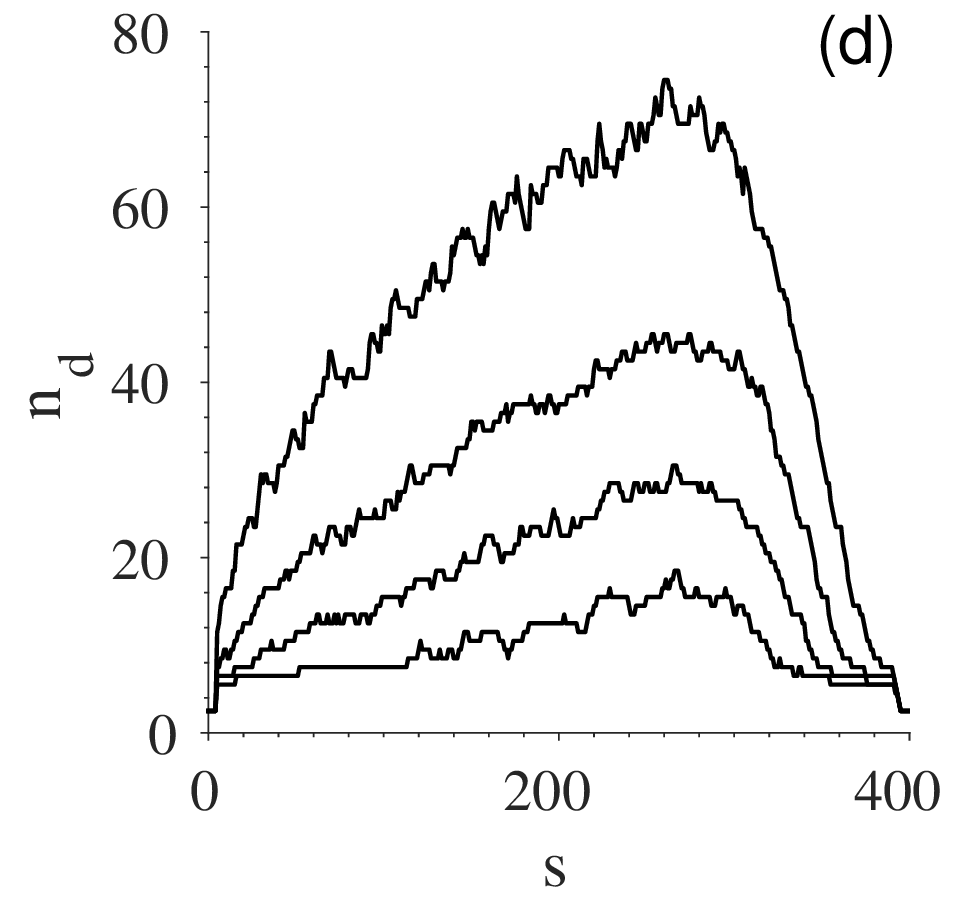}\\
\caption{Tension propagation in OTO (top) and  ATA (bottom) binding. Left column: Two-bond distances along the polymer around the $i^{th}$ PB as a function of the translocation coordinate $s$. $i=0$ labels the polymer end that translocates first. Darker shade of grey corresponds to larger distance. PBs on the \textit{cis} side are above the diagonal line and those on the \textit{trans} side are below it. The solid line above the diagonal corresponds to the two-bond distance $1.60$. Right column: The number of beads under drag. In each plot the curves from top to bottom correspond to different magnitudes of drag force with two-bond distance values starting from $1.60$ (top) and increasing by $0.02$ for each curve.}
\label{fig:TensionMatrixAndDrag} 
\end{figure}

Fig.~\ref{fig:TensionMatrixAndDrag} shows ensemble averages of the two-bond distances for polymers of length $N=400$ in translocations driven by OTO ((a) and (b)) and ATA ((c) and (d)) bindings. The tension propagation on the \textit{cis} side can be seen in the plots (a) and (c) as shaded areas above the diagonal. Tension propagation in the two models is clearly similar. In the ATA model the tension propagation is slightly more prominent as seen from the larger size and the darker shade of the area above the diagonal.

By extracting contours for different values of the two-point distance we gain a more precise picture of the tension propagation in different models. The number of beads $n_d$ experiencing a certain magnitude of drag can be calculated by subtracting the diagonal value from the value of $i$ for each $s$. The outcome is depicted in Figs.~\ref{fig:TensionMatrixAndDrag} (b) and (d). Shown are all two-bond distance values greater than the equilibrium value $1.59$ for our self-avoiding polymer. The top curve $n_d(s)$ in each subfigure corresponds to the contour for the two-bond distance value of $l_d = 1.60$. The subsquent $n_d(s)$ curves are plotted for $l_d = 1.62, 1.64,\ldots$ up to a value where the corresponding contour can no longer be distinguished from the diagonal of the respective left column plots of Fig.~\ref{fig:TensionMatrixAndDrag}. The higher the $l_d$ for the contours that can be distinguished is, the more prominent is the tension propagation. Hence, it can be seen that tension propagation is most prominent in the ATA binding. This can be accounted for by the ATA binding leading to faster translocation. 

\begin{figure}[t]
\includegraphics[width=0.49\linewidth]{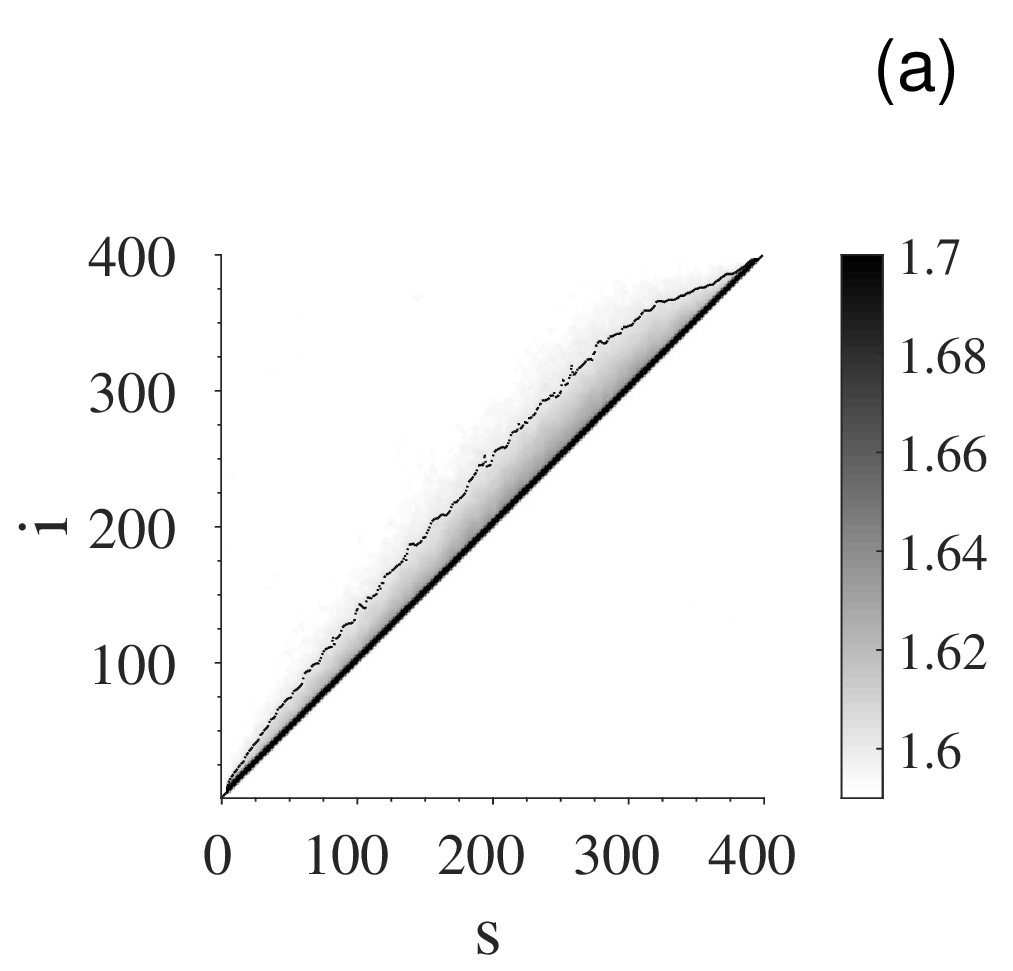}
\includegraphics[width=0.49\linewidth]{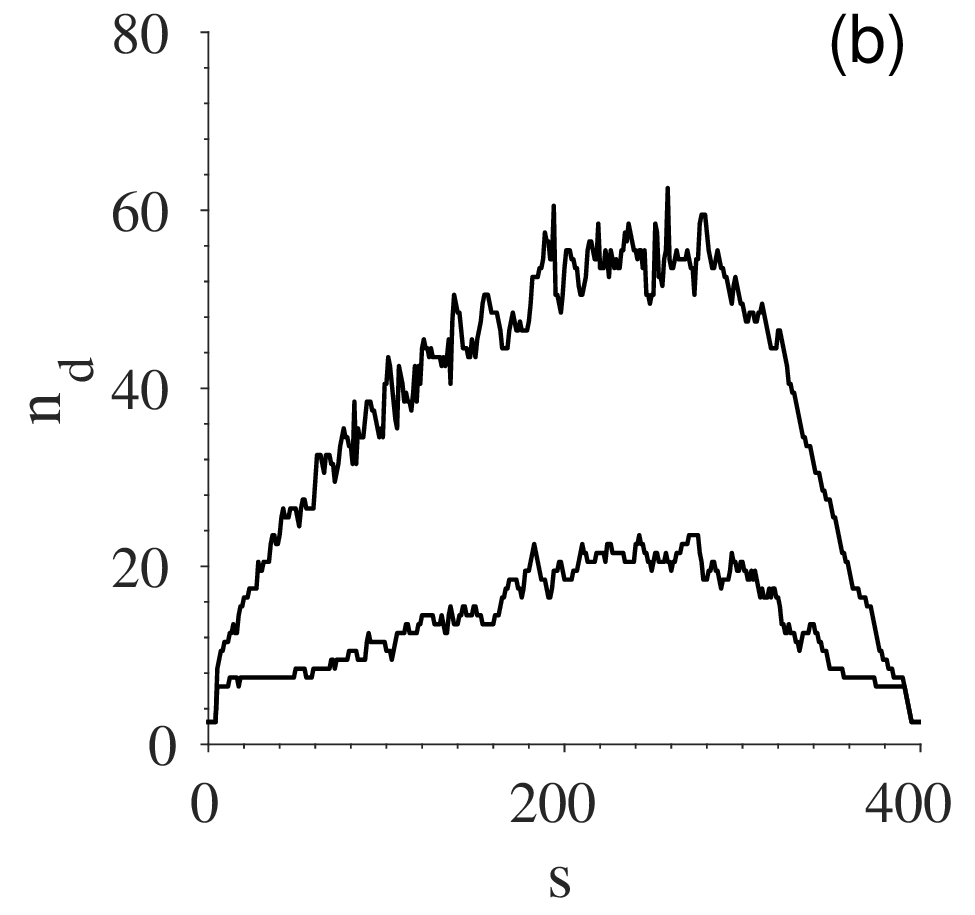}\\
\includegraphics[width=0.49\linewidth]{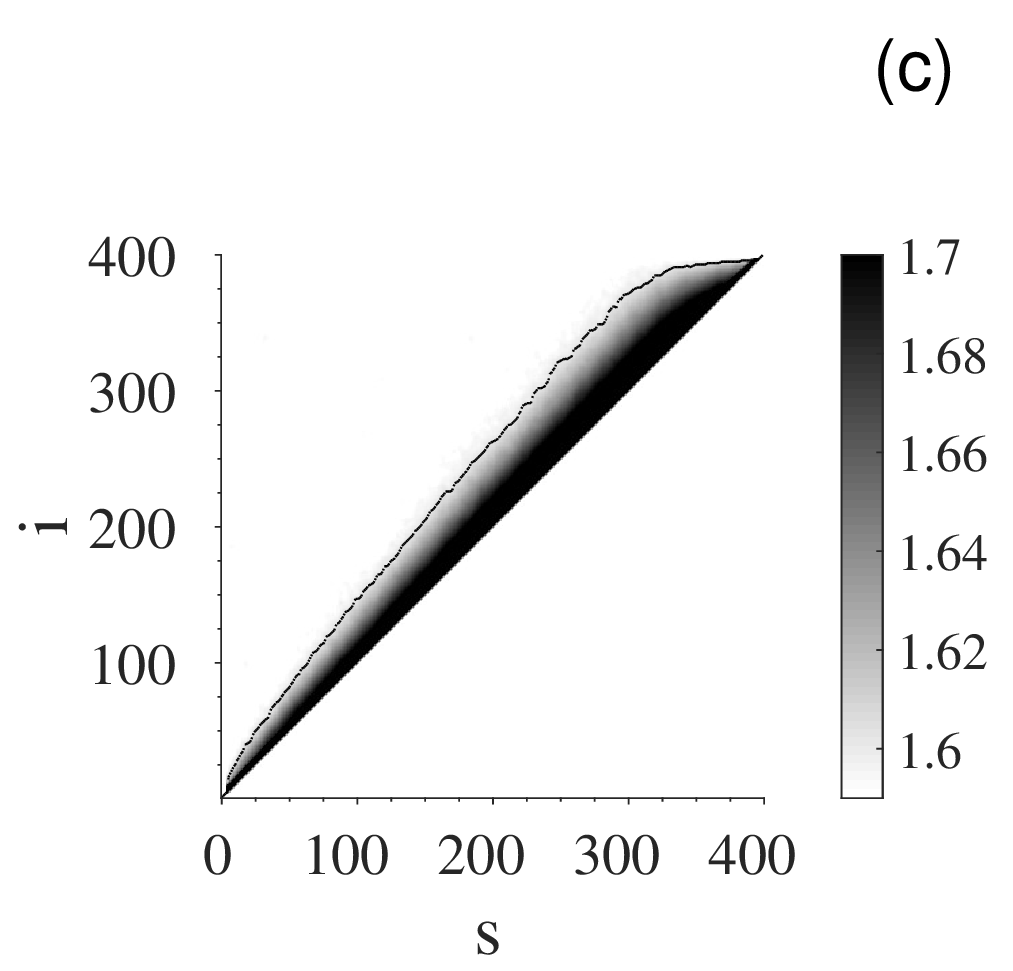}
\includegraphics[width=0.49\linewidth]{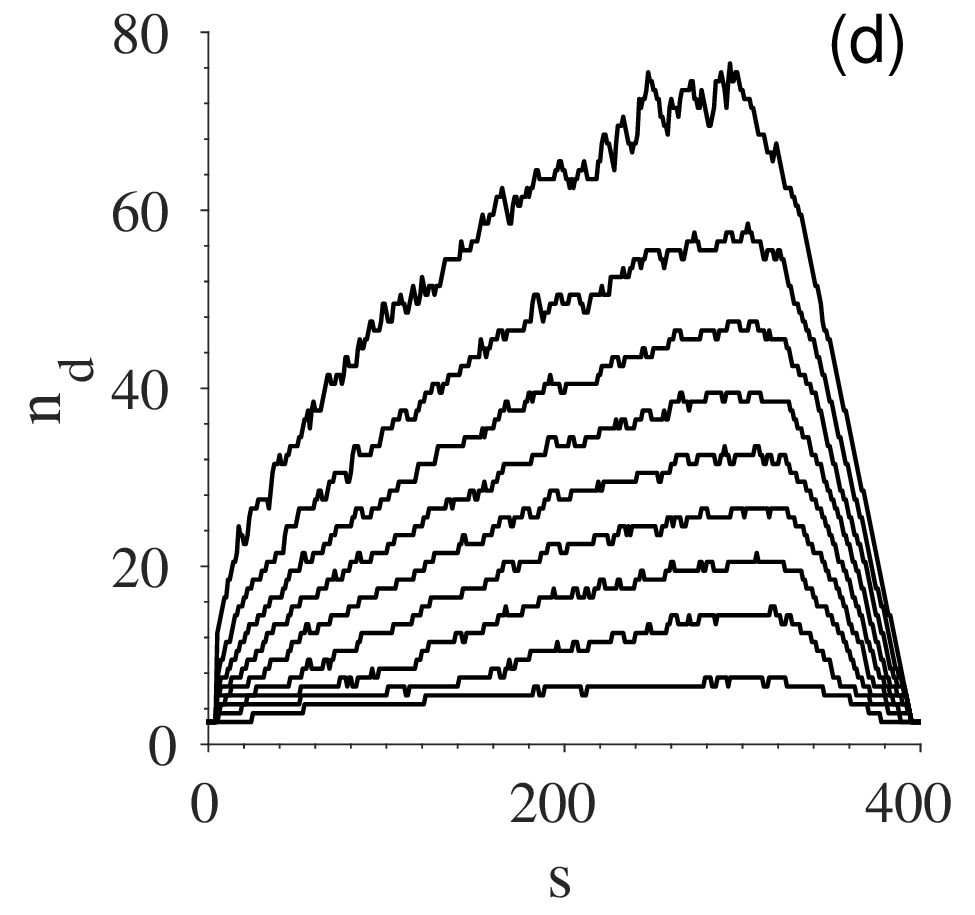}\\
\caption{From top to bottom: Driven translocation with a driving force $f_d=0.25$ and perfect Brownian ratchet. Left and right columns, respectively, present the corresponding data described in the caption of Fig.~\ref{fig:TensionMatrixAndDrag}}
\label{fig:TensionMatrixAndDrag2} 
\end{figure}

To further assess how largely tension propagation defines the translocation dynamics in the case of OTO binding we compare the waiting times and tension propagation in translocations driven by OTO and pore force.  We have previously shown that the {\it trans} side has no discernible contribution on the dynamics in the case of driven translocation~\cite{Suhonen14}. Hence, the  translocation driven by pore force can be used as a reference for polymer translocation whose dynamics is practically completely determined by tension propagation. Figs.~\ref{fig:TensionMatrixAndDrag2} (a) and (b) give the above-described tension propagation data for the driven polymer translocation. The pore force $f_d = 0.25$ was selected so that it takes the same average time for polymers of length $N = 400$ to complete the driven and the OTO translocation. Accordingly, the closest match of $t_w(s)$ is seen for $N = 400$.

\begin{figure}[!ht]
\includegraphics[width=.49\linewidth]{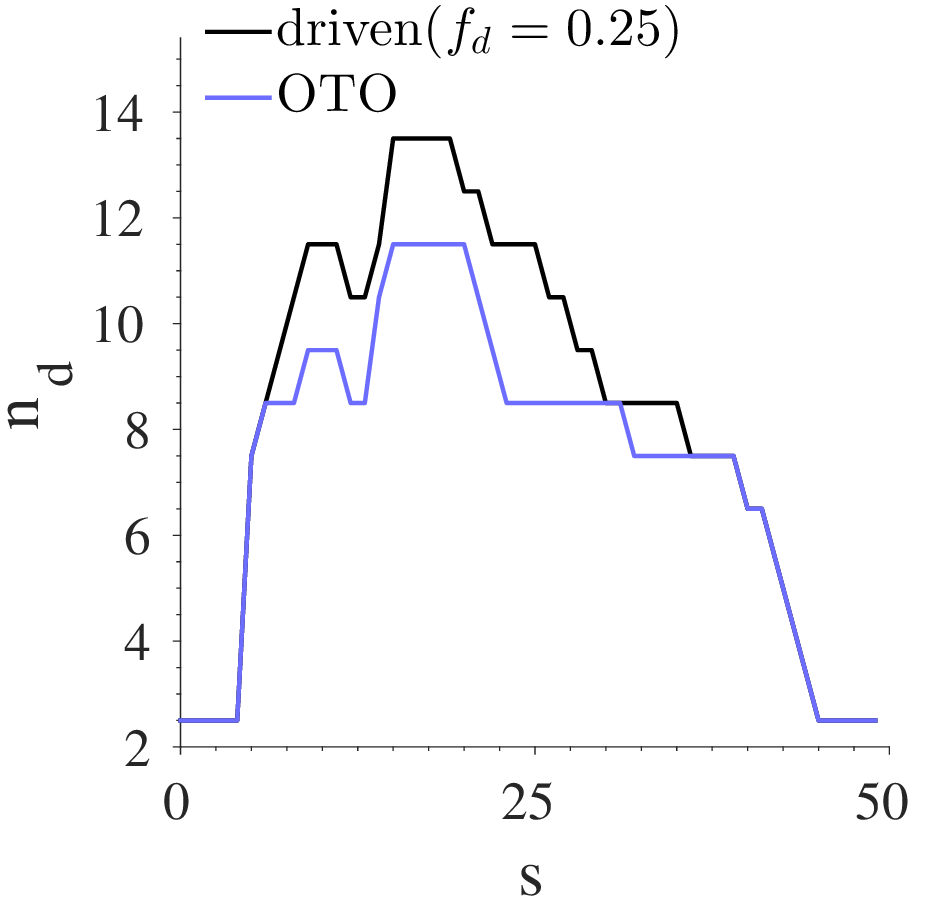}
\includegraphics[width=.49\linewidth]{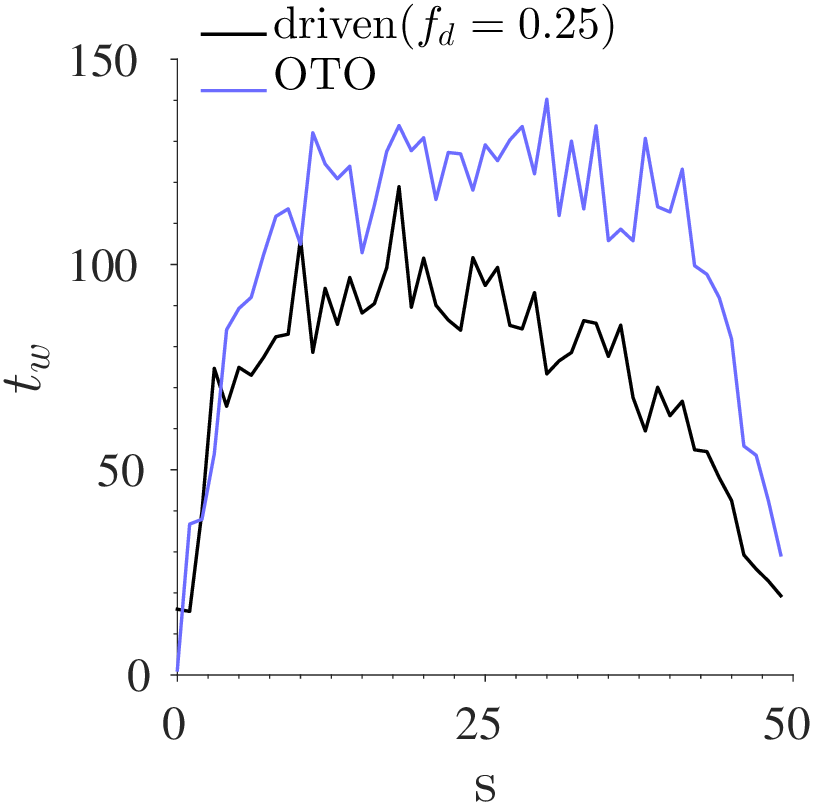}
\includegraphics[width=.49\linewidth]{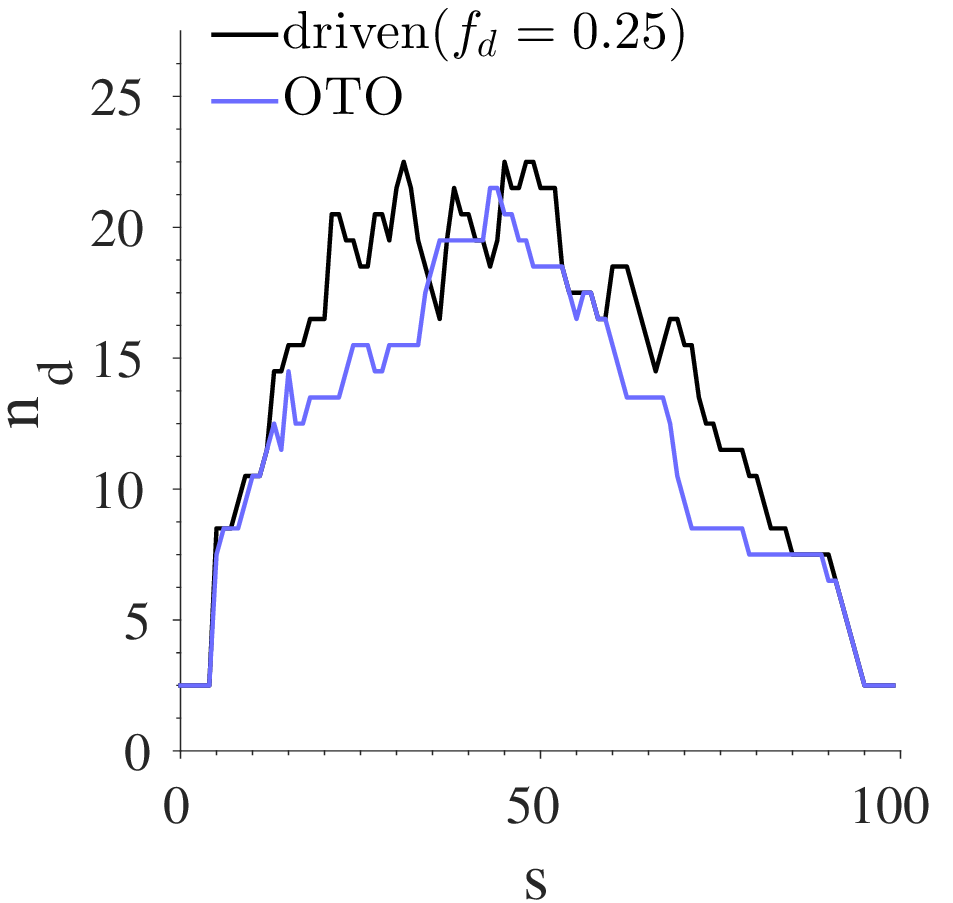}
\includegraphics[width=.49\linewidth]{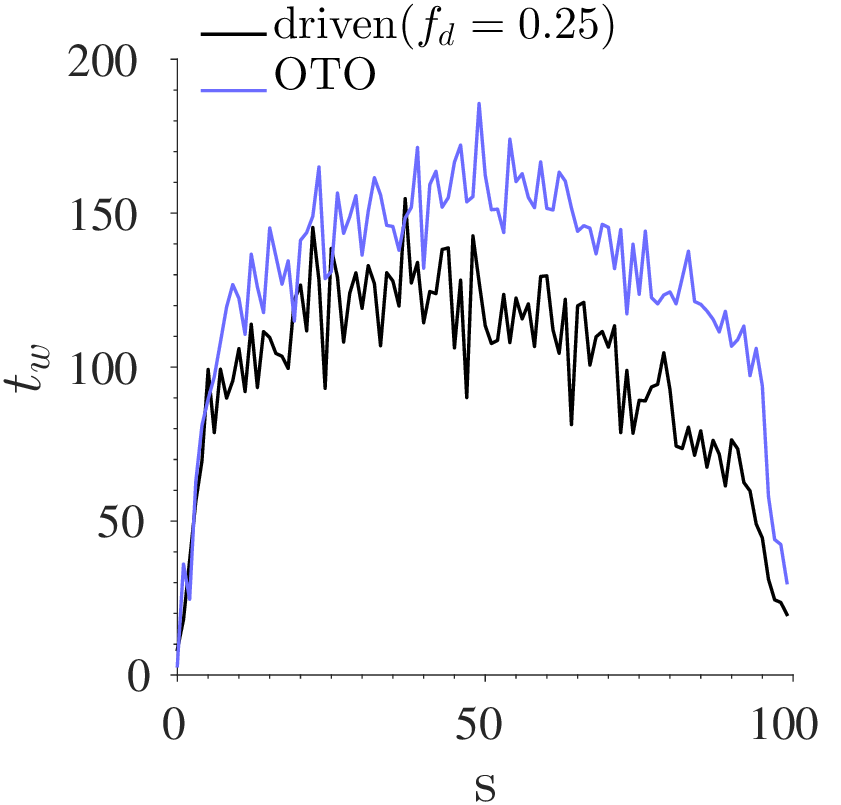}
\includegraphics[width=.49\linewidth]{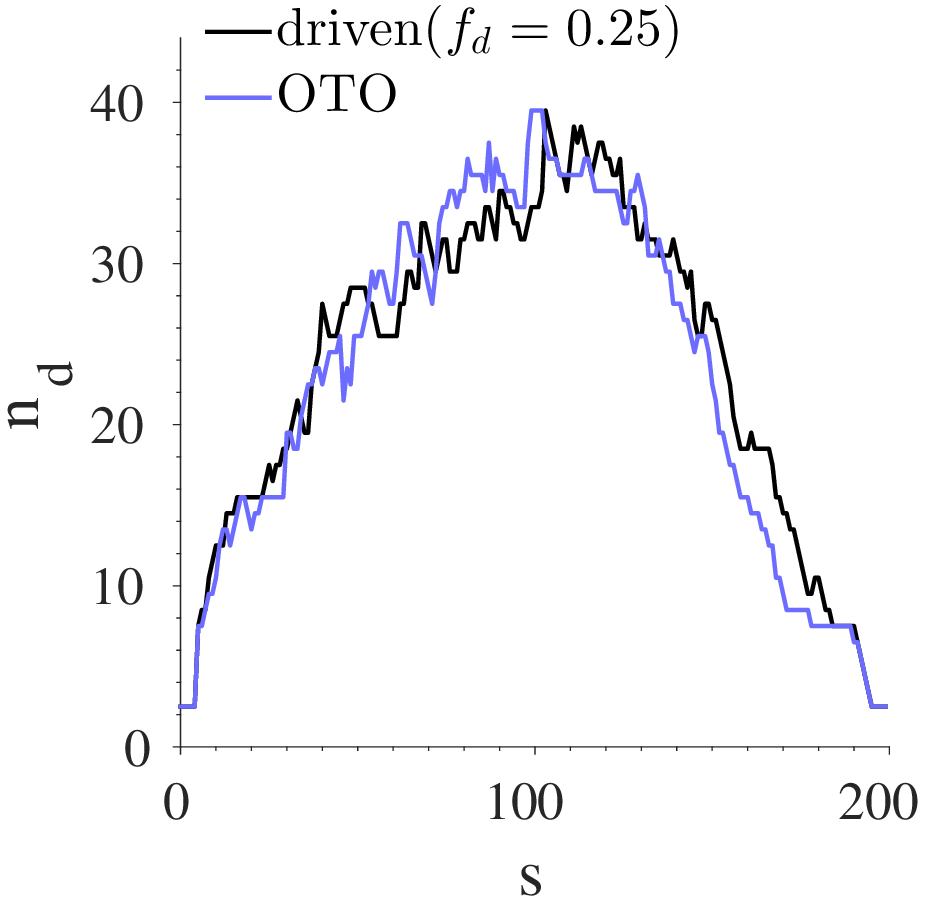}
\includegraphics[width=.49\linewidth]{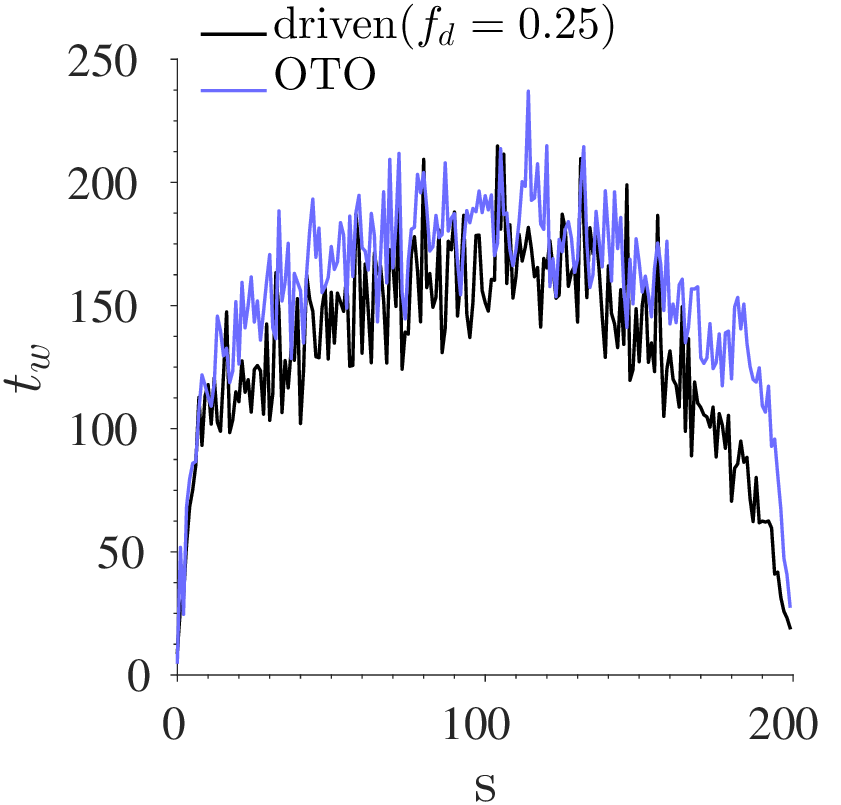}
\includegraphics[width=.49\linewidth]{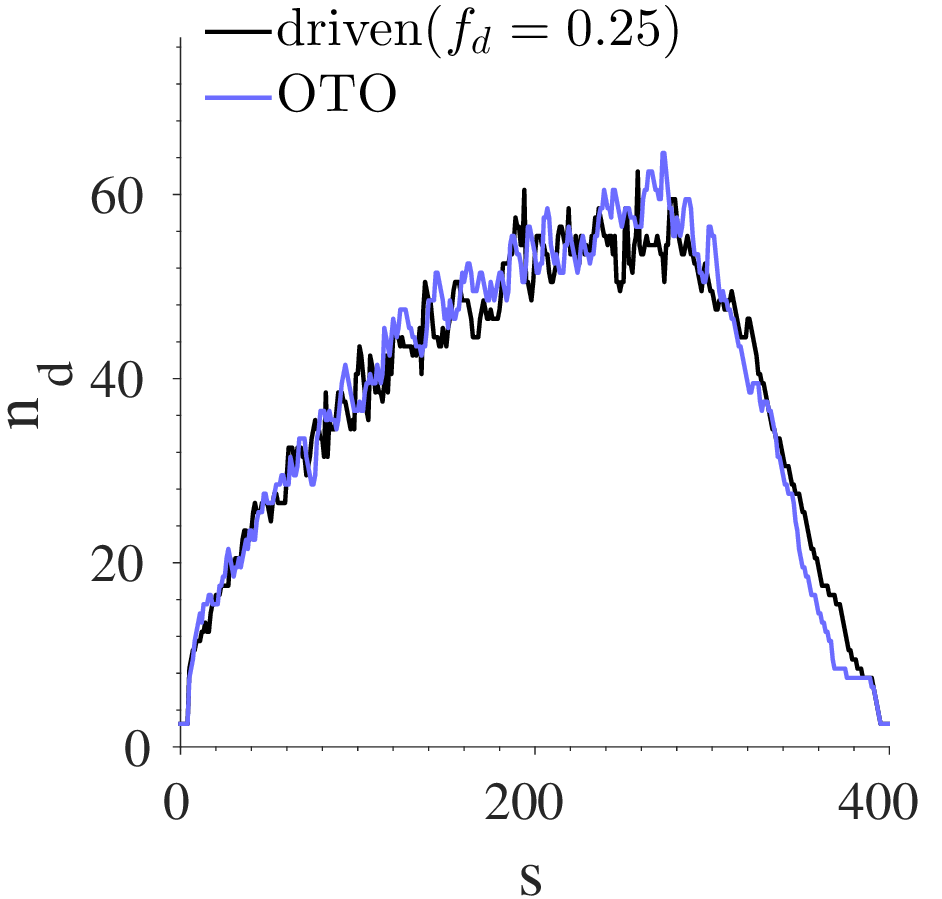}
\includegraphics[width=.49\linewidth]{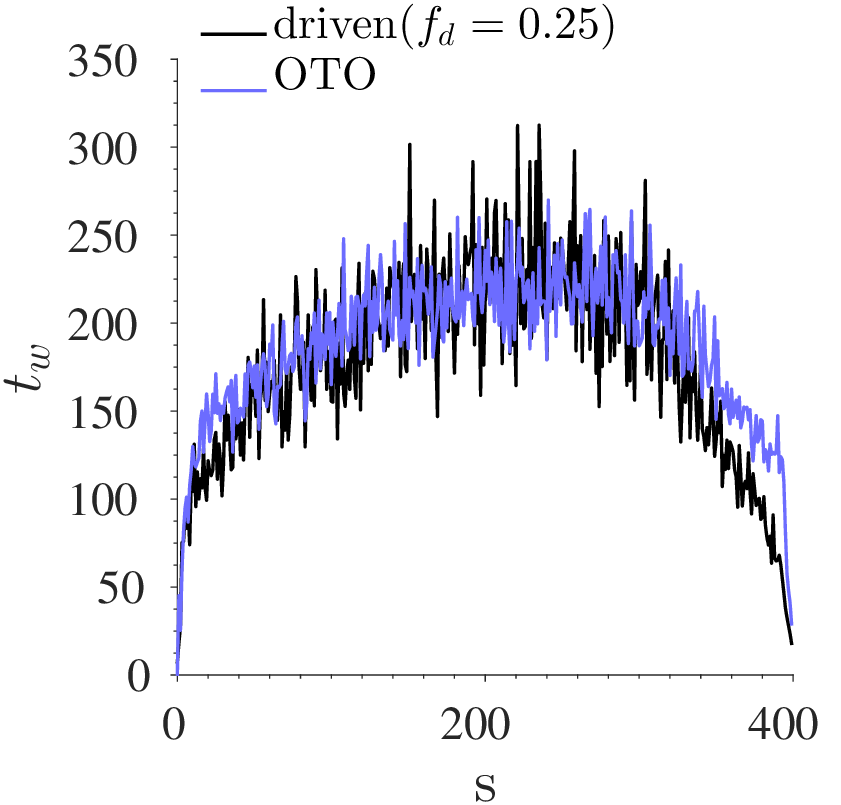}
\caption{(Color online) The number of beads in drag $n_d(s)$ (left column) and the waiting times $t_w(s)$ (right column) for OTO binding and driven tranloscation, $f_d = 0.25$.} 
\label{fig:OTOComb}
\end{figure}

Fig.~\ref{fig:OTOComb} compares OTO and $f_d$ driven translocation. Here, the extent of the tensed segment on the {\it cis} side in number of beads in drag $n_d$ is shown on the left and the waiting times $t_w$ as functions the number of translocated beads $s$ on the right column for different $N$. The tension on the {\it cis} side is seen to propagate identically in translocations driven by pore force and OTO binding. There are minor differences in the waiting time profiles. As the frictional contribution due to tension propagation on {\it cis} side is seen to be identical these differences have to come solely from the {\it trans} side where the binding changes the polymer conformation: the altered friction and inertia due to binding particles directly affect the translocation dynamics.

\subsection{Bias due to binding}

The bias driving the polymer through the pore is caused by two factors: energy drop on the {\it trans} side and Brownian ratcheting, both caused by the binding particles. For the completely stiff (rod) polymer it was found that the driving caused by the energy drop dominates over perfect Brownian ratchet mechanism~\cite{Zandi03}. To determine the dominating mechanism in the case of a flexible polymer we simulate a three-dimensional translocation model where the polymer is driven by perfect ratcheting only. The model geometry is the same as in ATA-, OTO-, and $f_d$-driven models. There is no driving force nor binding particles, only the backward motion of the polymer segment inside the pore is completely inhibited to realize perfect ratcheting. Fig.~\ref{fig:TensionMatrixAndDrag2} (c) and (d) show the tension propagation characteristics for perfect Brownian ratcheting. Tension propagation for the perfect Brownian ratchet is seen to be clearly the strongest of the different models.

\begin{figure}[t]
\includegraphics[width=0.49\linewidth]{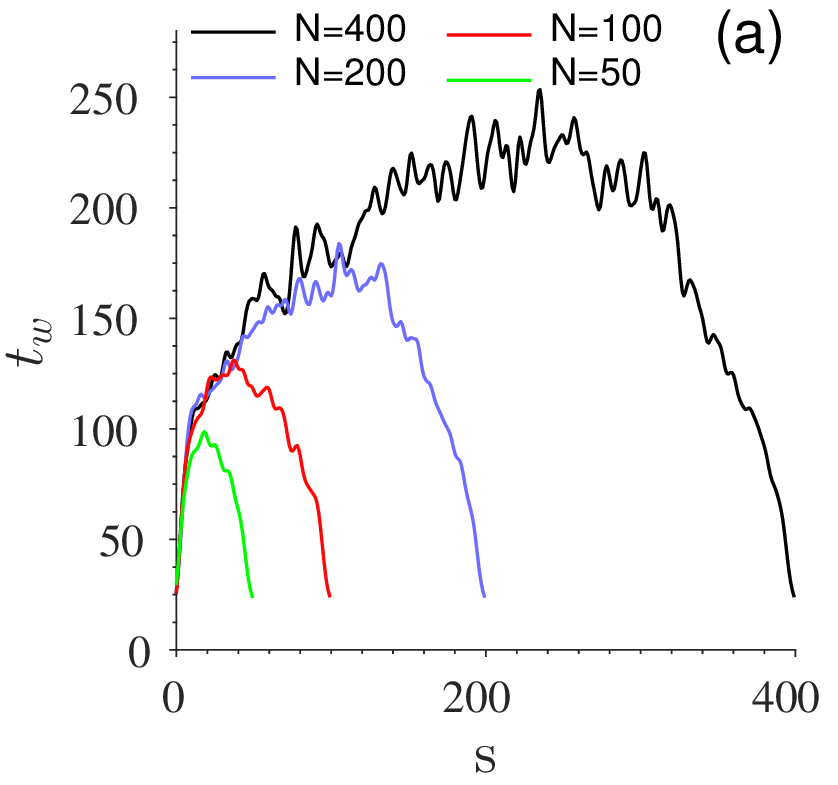}
\includegraphics[width=0.49\linewidth]{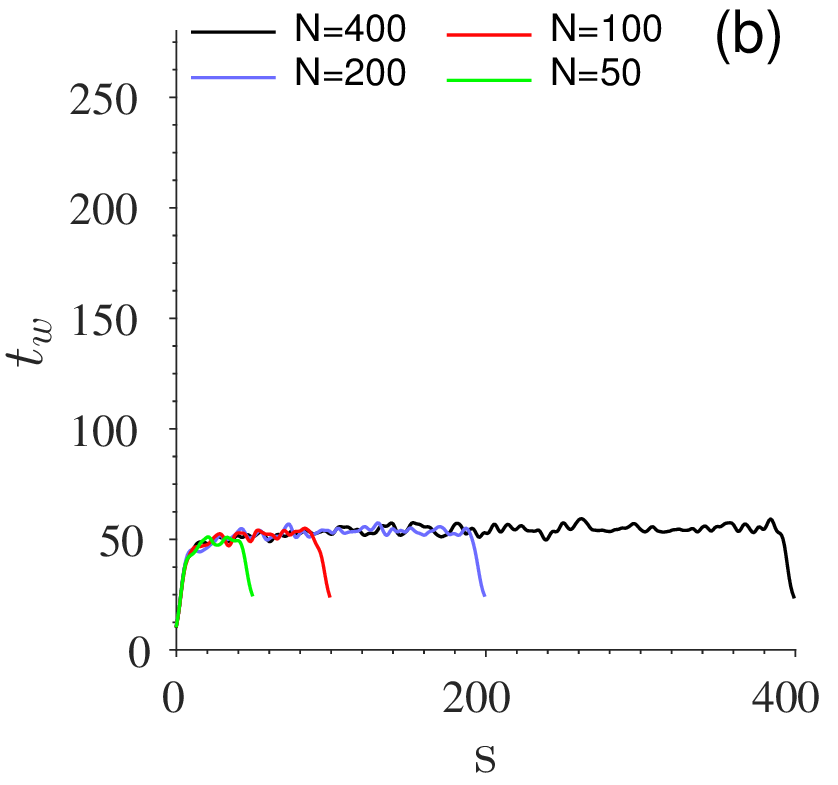}
\includegraphics[width=0.49\linewidth]{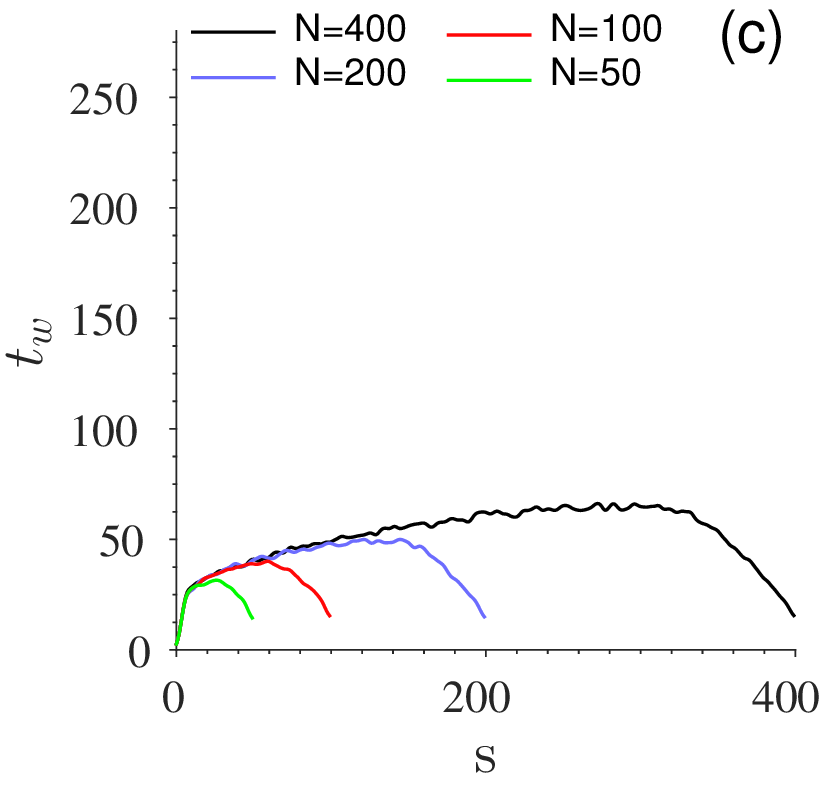}
\includegraphics[width=0.49\linewidth]{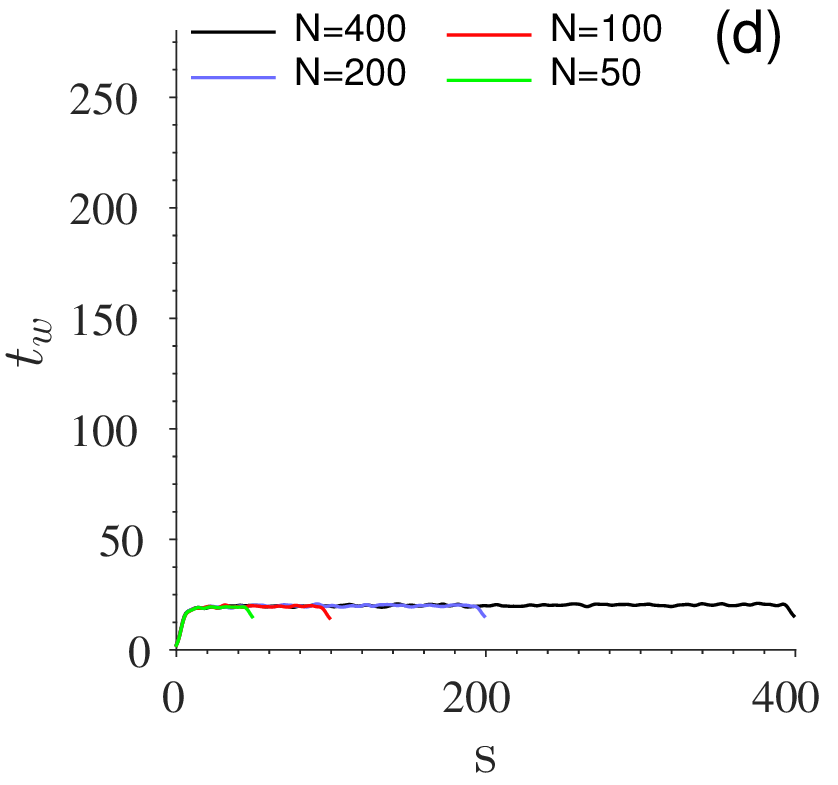}
\caption{(Color online) Waiting times for full translocation models (left column) and models where the contribution from the {\it cis} side is excluded (right column). The first row: the  driven translocation model, $f_d=0.25$. The second row: the perfect ratchet translocation model. Polymer lengths $N=50$, $100$, $200$, and $400$.\color{black}}
\label{fig:WaitingTimesRef}
\end{figure}

In Fig.~\ref{fig:WaitingTimesRef} $t_w(s)$ for the full models and ones where {\it cis} side is excluded are given for the driven translocation and the perfect ratchet model. The perfect Brownian ratchet mechanism is seen to be clearly faster than the translocations driven by constant force and the binding particles, see Fig.~\ref{fig:WaitingTimes}. As seen in Fig.~\ref{fig:WaitingTimesRef}, eliminating the {\it cis} side in Brownian ratchet dynamics results in a completely flat $t_w(s)$. For the full ratchet model $t_w(s)$ is identical in form to that of the driven translocation, see  Fig.~\ref{fig:WaitingTimesRef}. This confirms that tension propagation on the {\it cis} side predominantly determines the dynamics in perfect Brownian ratcheting like in the driven translocation.

In the OTO model particle unbinding allows for some backward motion of the polymer, so the ratcheting mechanism is not perfect. Our simulations show that the model with Brownian ratcheting alone without energy reduction due to binding gives by far the strongest bias of all the simulated modes. This suggests that it is the Brownian ratcheting that dominates in $3$ dimensions the translocation of a flexible polymer by binding particles, not the reduction of the free energy on the {\it trans} side due to binding. This is in contrast what was found for chaperone-assisted translocation of stiff polymers~\cite{Zandi03}.

\subsection{Translocation time vs. polymer length}\label{sec:scaling}

\begin{figure}[t]
\includegraphics[width=1.00\linewidth]{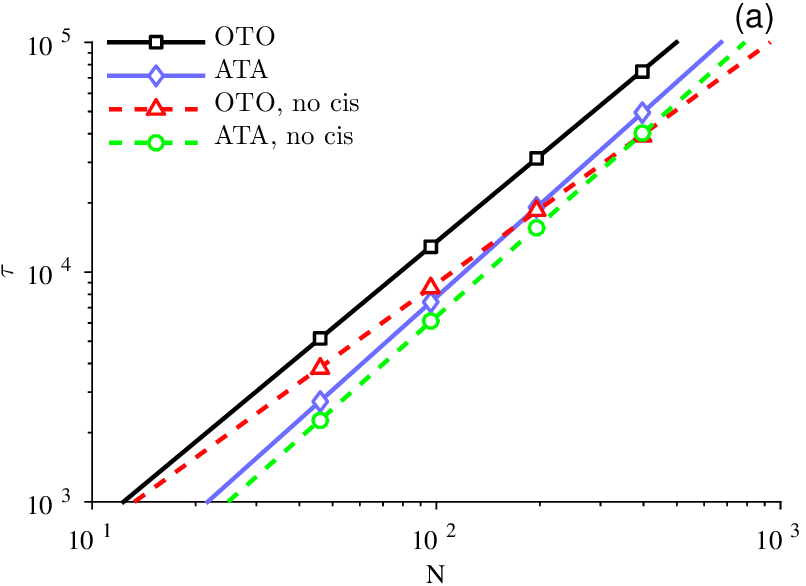}\\
\includegraphics[width=1.00\linewidth]{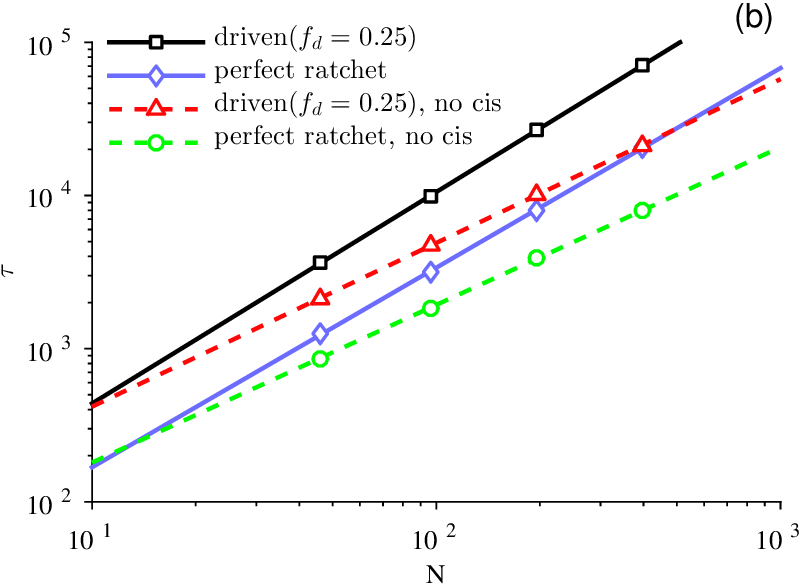}
\caption{(Color online) Scaling of the translocation times. Scaling exponents obtained by fitting  $\tau \sim N^\beta$ to the data. (a) The BiP driven models. OTO: $\beta=1.26 \pm 0.02$ , ATA: $\beta=1.36 \pm 0.01$, OTO no \textit{cis}: $\beta=1.09 \pm 0.01$, and , ATA no \textit{cis}: $\beta=1.34 \pm 0.02$. (b) The reference models. Translocation driven by the pore force $f_d=0.25$: $\beta=1.39 \pm 0.02$, by perfect Brownian ratchet mechanism: $\beta=1.32 \pm 0.01$, driven($f_d=0.25$) no \textit{cis}: $\beta=1.08 \pm 0.02$, and perfect ratchet no \textit{cis}: $\beta=1.04 \pm 0.01$\color{black}}
\label{fig:LengthScaling}
\end{figure}

Here, we verify the above-presented analysis by looking at the scaling of translocation time $\tau$ with polymer length $N$ in the different model systems. Fig.~\ref{fig:LengthScaling} shows average $\tau$ as a function of $N$ for the full and modified binding models. The error bars of the data points are much smaller than the used symbols. Fig.~\ref{fig:LengthScaling} also shows the scaling relations $\tau \sim N^\beta$ fitted to the data. The scaling exponent are $\beta=1.26 \pm 0.02$ and $\beta=1.09 \pm 0.01$ for the full OTO model and one where the polymer segment on the \textit{cis} side is excluded, respectively. The corresponding exponents for the full and modified ATA models are $\beta = 1.36 \pm 0.01$ and $1.34 \pm 0.02$, respectively.

Removal of segments on the \textit{cis} side reduces $\beta$ from $1.26$ to $1.09$ in the OTO binding. The drop of $\beta$ to almost $1.0$, {\it i.e.} linear scaling confirms our observation that the \textit{trans} side has only a minimal effect on the translocation driven by OTO binding and, consequently, the weak tension propagation on the \textit{cis} side largely determines the dynamics. The fairly low value of $\beta=1.26$ is understandable, since particles binding to the polymer in the vicinity of the pore increase the local friction there. This leads to reduced $\beta$ for polymers of modest length~\cite{Lehtola09}.

The obtained superlinear scaling with $\beta > 1$ due to the \textit{trans} side could potentially come from crowding of the segment close to the pore. In the driven translocation the effect of the crowding was shown to be negligible. Inclusion of the {\it trans} side was nevertheless found essential as only then $\beta$ increased with $f_d$~\cite{Suhonen14}. This was addressed to fluctuations assisting translocation~\cite{Dubbeldam13}. However, the driving bias due to chaperones is weaker and small perturbations on the {\it trans} side are expected to show more easily in the outcome. Moreover, unlike in $f_d$ driven translocation crowding may play a role in the BiP driven case, since BiPs increase the time it takes for the translocated segments to relax to thermal equilibrium.  The binding rate may also slow down due to the diffusion of the binding particles toward the pore changing as the polymer translocates. This would slightly diminish the driving bias. Both these effects increase $\beta$.

In accordance with the observations from the waiting time profiles, removal of segments on the \textit{cis} side has only a small effect on the translocation driven by ATA binding, see Fig.~\ref{fig:LengthScaling}, which confirms that in this model the dynamics is almost solely determined by the translocated and collapsed polymer segment on the {\it trans} side. The correlation length of this densely crowded segment is very high. Accordingly, the collective motion of the segment is expected to be more important than the motion of individual monomers. Also, driving due to ATA binding is strong. If the {\it cis} side played a dominant role in the dynamics, then in the theoretical limit of extremely strong driving where the polymer segment would be instantly drawn from the {\it cis} side to the pore and $\beta \to 1 + \nu \approx 1.6$. However, the measured $\beta = 1.36$ obtained for $N \le 400$ is far below this and, as shown, comes mainly from the {\it trans} side.

The measured value for $\beta$ in ATA binding can be understood as follows. For the moment, we assume that the number of binding close to the pore, which determines the driving force, is approximately constant. Based on our measurements of the binding and unbinding during translocation this is not far from the truth. Consequently, in this approximation the bias due to binding and hence the momentum in the direction of translocation $\mathbf{p} = p = m v$,  are constant. Here, $\mathbf{v} = v$ is the (scalar) translocation velocity in the direction perpendicular to the wall and $m$ is the moving mass. Due to the strong attraction between monomers where BiPs attach, we assume the average distance from the pore to which the center-of-mass point has been moved on the {\it trans} side to scale with the number of translocated monomers as the gyration radius of the expanding globular conformation $\langle d \rangle \sim \langle R_g \rangle \sim s^{1/3}$. The mass of the packed globule on the {\it trans} side grows as $m \sim s$, which leads to $p \sim s v$ and, consequently, $v \sim 1/s$. The time-average over the whole translocation scales like $v$: $v = v_{\tau} =  1/\tau \int_0^{\tau} v dt = 1/\tau \int_0^{N} v(s) \frac{dt}{ds} ds \sim N^{-1}$. Accordingly, it can be taken as the effective velocity over the whole process, $\langle v \rangle = \langle v_{\tau} \rangle$. The average translocation time, as $s \to N$, then becomes $\tau = \langle d \rangle / \langle v \rangle \sim N^{4/3}$.

In reality, the effective bias due to binding of course varies somewhat, due to which $p$ does not remain strictly constant. Also $R_g$ does not scale strictly spherically. Departure from these assumptions cause $\beta$ to deviate from the predicted value $\beta = 4/3$. Still, the measured value $\beta = 1.36$ is very close.

In translocations driven by perfect Brownian ratchet mechanism scaling relations for the full model and one where polymer segments on the {\it cis} side are removed confirms that the dynamics is mainly determined by the tension propagation on the {\it cis} side, see Figs.~\ref{fig:WaitingTimesRef}~(c) and (d). From Fig.~\ref{fig:LengthScaling} (b) the scaling exponents $\beta$ are seen to be somewhat smaller than for the translocation driven by a constant pore force $f_d$. $R_g(s)$ on the {\it trans} side for the perfect ratchet model and translocation driven by $f_d$ are almost identical (not shown), so based on our previous results~\cite{Suhonen14} in spite of $R_g(s)$ being smaller than the equilibrium $R_g$ this crowding on the {\it trans} side has no effect on translocation dynamics.

As described, our perfect ratchet model does not involve any binding particles but ratcheting comes from not allowing the polymer to slide back toward {\it cis}. Hence, the only qualitative difference to constant-force-driven translocation is that fluctuations in reaction coordinate $s$ are rectified. In other words, fluctuations that would move the polymer back toward {\it cis} are eliminated and only forward directed fluctuations are allowed. Hence,  the assistance of the fluctuations in translocation is further enhanced compared to driven translocation~\cite{Dubbeldam13}, resulting in a smaller $\beta$.

\subsection{Concentration and binding force dependence of the translocation time}\label{sec:coneteps}

In previous sections the free BiP concentration and the binding constant were set at $c_f = 1/40$ and $\epsilon_b = 8.0$, respectively. Here we investigate how the translocation times are affected when $c_f$ is varied between $1/320$ and $1/5$ and binding strength $\epsilon_b$ between $1.0$ and $64.0$.

\begin{figure}[t]
\includegraphics[width=1.0\linewidth]{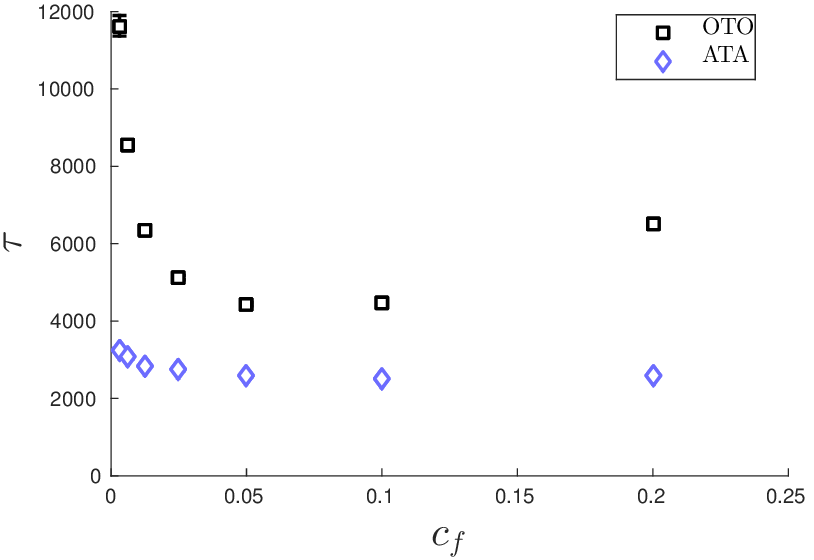}
\caption{(Color online) Average translocation times $\tau$ as a function of the free BiP concentration $c_f$ when $\epsilon_b=8$ and $N=50$.}
\label{fig:TransTimeChapCon}
\end{figure}

Fig.~\ref{fig:TransTimeChapCon} shows the average translocation times $\tau$ as a function of $c_f$. The simulations were done for $N=50$ and $\epsilon_b=8$. It can be seen that for the OTO there is a clear minimum of translocation speed as a function of $c_f$. This is in accord with the results for the translocation driven by ATA binding in two dimensions~\cite{Yu11,Yu12}. There the increase of $\tau$ after initial decrease when increasing $c_f$ was related to additional friction due to binding BiPs and also the running out of BiPs, since a constant number of BiPs was used. In our simulations concentration of {\it free} binding particles is kept constant, so BiPs do not run out, and the contribution that remains is the increased friction.

For ATA the translocation times decrease with increasing $c$ and reach a minimum without increasing again. Hence, translocation by ATA binding in three dimensions differs from that in two dimensions~\cite{Yu11,Yu12}. Due to smaller spatial restrictions intersegmental binding in three dimensions is much more pronounced, so driving due to this binding is not inhibited when increasing $c_f$ in the same way as in two dimensions. 

When increasing the binding constant $\epsilon_b$ in our simulations $\tau$ rapidly decreases saturating to a constant minimum value for both binding models (not shown). This indicates that no spatial restrictions emerge for binding in three dimension for these $c_f$ and $\epsilon_b$, which in part supports the approximation of constant bias made in deriving the scaling $\tau \sim N^\beta$ for translocation driven by ATA binding.

\section{Conclusion}\label{sec:con}

We have studied chaperone-assisted translocation of flexible polymers through a nanometer-scale pore in three dimensions by computer simulations using models based on Langevin dynamics. We implemented two mechanisms for the chaperones to bind to the polymer on the {\it trans} side. In one-to-one (OTO) binding a chaperone can bind to only one site, whereas in all-to-all (ATA) binding it can bind to multiple sites on the polymer simultaneously. We showed that in three dimensions the differences in binding lead to substantial differences in translocation dynamics.

In the OTO binding the polymer is driven increasingly out of equilibrium much the same way as in the case of constant pore force $f_d$ driving the translocating polymer. We showed that for this binding tension propagates on the {\it cis} side in exactly the same way as in $f_d$-driven translocation. In spite of this similarity waiting time profiles showed differences for the two cases. Translocation assisted by OTO binding is slightly slowed down compared to the $f_d$-driven case. Obviously, the differences have to come from the {\it trans} side. Crowding of the polymer segment, which we have previously shown not to affect $f_d$-driven translocation, can to some extent impede chaperone-assisted translocation, since the inertia and friction of the polymer segment on the {\it trans} side is increased due to binding particles.

The main conclusion concerning OTO-binding assisted translocation in three dimensions is that its dynamics is mainly determined by tension propagation on the {\it cis} side and that the tension propagates exactly like the tension in translocation driven by pore force whose magnitude equals the bias due to binding chaperones. The exponent for scaling of the translocation time with the polymer length, $\tau \sim N^\beta$, in OTO-binding assisted translocation was found to be $\beta \approx 1.26$.  This value is low given the similarity of the process to the pore-force driven case. One explanation for this is the increased local friction due to chaperones binding in the vicinity of the pore on the {\it trans} side~\cite{Lehtola09}.

Under the ATA binding the polymer conformation on the {\it trans} side is very dense and accordingly motion of the monomers in it is highly correlated. We found that although tension propagation on the {\it cis} side is strong due to rapid translocation, contribution of the {\it trans} side dominates the dynamics. We derived the scaling exponent $\beta = 4/3$ for the approximated case of a completely correlated moving (and growing) spherical polymer conformation on the {\it trans} side under constant bias translocation. This is very close to the value $\beta \approx 1.36$ obtained from our simulations.

To summarize, chaperone-assisted translocation of flexible polymers in three dimensions is highly dependent on the binding mechanism. Clear similarity to translocation driven by constant pore force was found for the single-binding scenario, whereas allowing binding to take place on multiple sites simultaneously changed the picture dramatically. The results presented here will pave the way for detailed understanding and possibly application of the many variations of chaperone-assisted biopolymer translocation.

\begin{acknowledgments}
We thank prof.~Jouko Lampinen, Dept. of Computer Science, Aalto University, for his support to this work. The computational resources of CSC-IT Centre for Science, Finland, and Aalto Science-IT project are acknowledged.
\end{acknowledgments}

\bibliography{references.bib}

\end{document}

%% file: copyrightaps.tex
{
\thispagestyle{empty}
\addtocounter{page}{-1}
\onecolumngrid
\raggedright{}
\Huge 
Copyright Notice\\
\vspace{10mm}
\normalsize
Copyright (2016) by the American Physical Society.\\
\vspace{10mm}
Chaperone-assisted translocation of flexible polymers in three dimensions\\
P. M. Suhonen, and R. P. Linna\\
\vspace{10mm}

Citation: Phys. Rev. E 93, 012406 (2016)\\
URL: \href{https://journals.aps.org/pre/abstract/10.1103/PhysRevE.93.012406}{https://journals.aps.org/pre/abstract/10.1103/PhysRevE.93.012406}\\
DOI: 10.1103/PhysRevE.93.012406\\
\clearpage
}